\documentclass[%
 reprint,
 amsmath,amssymb,
 aps,
]{revtex4-2}

\usepackage{graphicx}
\usepackage{dcolumn}
\usepackage{bm}
\usepackage{hyperref}
\usepackage[mathlines]{lineno}

\begin{document}

\preprint{INT-PUB-25-001}

\title{Description of di-hadron saturation signals within a universal nuclear parton distribution function approach}

\author{Dennis V. Perepelitsa}
\email{dvp@colorado.edu}
\affiliation{%
 University of Colorado Boulder \\
 Boulder, CO 80309
}%


\begin{abstract}
Di-hadron and di-jet correlation measurements in proton--nucleus ($p$+A) and electron--nucleus collisions are widely motivated as sensitive probes of novel, non-linear QCD saturation dynamics in hadrons, which are particularly accessible in the dense nuclear environment at low values of Bjorken-$x$ ($x_\mathrm{A})$.
Current measurements at RHIC and the LHC observe a significant suppression in the per-trigger yield at forward rapidities compared to that in proton--proton collisions, nominally consistent with the ``mono-jet'' production expected in a saturation scenario. 
However, the width of the azimuthal correlation remains unmodified, in contradiction to the qualitative expectations from this physics picture. 
I investigate whether the construction of these observables leaves them sensitive to effects from simple nuclear shadowing as captured by, for example, universal nuclear parton distribution function (nPDF) analyses.
I find that modern nPDF sets, informed by recent precision measurements sensitive to the shadowing of low-$x_\mathrm{A}$ gluon densities in LHC and other data, can describe all or the majority of the di-hadron/jet suppression effects in $p$+A data at both RHIC and the LHC, while giving a natural explanation for why the azimuthal correlation width is unmodified.
Notably, this is achieved via a $(x_\mathrm{A},Q^2)$-differential suppression of overall cross-sections only, without requiring additional physics dynamics which alter the inter-event correlations.
\end{abstract}

\maketitle


\section{Introduction}

Proton--nucleus ($p$+A) collisions at the Relativistic Heavy Ion Collider (RHIC) and the Large Hadron Collider (LHC) serve as experimental opportunities to probe novel parton dynamics inside high-density nuclei~\cite{Salgado:2011wc,Salgado:2016jws}. 
Multiple theoretical approaches attempt to describe hard and semi-hard scattering processes in $p$+A collisions. One approach is based on leading-twist perturbative QCD in a collinear factorization picture, where all initial and final state effects on hard process rates are contained within a set of empirically-determined universal nuclear parton density functions (nPDFs)~\cite{Ethier:2020way}. The nPDFs are taken to be process-independent and depend only on the Bjorken-$x$ in the nucleus ($x_\mathrm{A}$) and the parton momentum transfer $Q^2$. Calculations based on the nPDF picture have been successful in describing how jet, hadron, and electroweak boson cross-sections (e.g.) are modified in collisions involving nuclei over a very large kinematic range. Other theoretical approaches are based on first-principles, dynamical descriptions of the initial state of the cold nucleus, such as from coherent multiple scattering of the partons participating in the hard scattering~\cite{Qiu:2003vd,Qiu:2004da} or within the Color Glass Condensate (CGC) effective theory framework~\cite{McLerran:1993ni,McLerran:1993ka,Gelis:2010nm,Morreale:2021pnn}. The quite different nature of these approaches raises an important question as to what extent these paradigms are distinct and what range of phenomena each should be expected to describe.

A key topic in contemporary heavy-ion and future Electron Ion Collider (EIC) physics is the search for definitive evidence of the onset of gluon saturation in collisions involving nuclei. These should give rise to a variety of novel effects besides just the suppression of cross-sections from nuclear shadowing, for example as encoded in the nPDFs. In the theoretical literature, di-hadron or di-jet correlation measurements in $p$+A collisions have long been motivated as a way to identify the onset of these non-linear QCD effects, particularly within the CGC theory~\cite{Kharzeev:2004bw,Jalilian-Marian:2004vhw,Marquet:2007vb,Stasto:2011ru,Kutak:2012rf,Albacete:2018ruq,Fujii:2020bkl}. Schematically, the incoming parton in the proton interacts coherently with a saturated gluon field in the nucleus, scattering to produce a ``mono-jet'' with no recoiling partner above some kinematic threshold. The increased prevalence of such mono-jet configurations manifests as a decreased yield of ``associated'' hadrons or jets, conditional on the presence of a ``trigger'' hadron or jet, and a broadening of the azimuthal correlation between detected pairs.
In this picture, one generally expects both suppression and broadening effects to appear together, although the specific degree of broadening may be sensitive to the inclusion of various NLO effects~\cite{Caucal:2023fsf}.
These phenomena are most relevant at forward (proton-going) rapidities, which select processes with a low $x_\mathrm{A}$. 
Importantly, future di-hadron correlation measurements in electron--nucleus ($e$+A) collisions have also been identified as one of the essential signals of saturation at the EIC~\cite{Accardi:2012qut,Zheng:2014vka,Caucal:2023fsf}. 

Early measurements of forward di-hadron correlations in deuteron--gold ($d$+Au) collisions have been performed by STAR~\cite{STAR:2006dgg,Braidot:2010zh} and PHENIX~\cite{PHENIX:2011puq} at RHIC, showing a significant suppression of the per-trigger yield compared to that in $p$+$p$ collisions in the same kinematics. At the time, multiple authors considered the question of whether a picture based on perturbative QCD with the contemporary set of collinear nPDFs (i.e. and no other dynamics) can describe such signatures, e.g. Refs.~\cite{Albacete:2010rh,Chiu:2011ya,ChuTalk}.
This is challenging because an nPDF-based picture has only one mechanism to affect such observables, which is via a reweighting of cross-sections in a way that is universal in $(x_\mathrm{A}, Q^2)$, and it cannot otherwise modify particular properties of events such as the kinematics of final states, their correlations, etc., outside of this prescription. Thus, measuring the suppression and broadening observables together is necessary to discriminate between the physical origin of effects. Approaches based on perturbative QCD calculations of nuclear effects~\cite{Kang:2011bp} have also been proposed as competing explanations to the saturation picture.

Recently, global nPDF analyses have begun to include new datasets from high-precision measurements at the LHC~\cite{Armesto:2015lrg,Eskola:2019dui,AbdulKhalek:2020yuc,Kusina:2020lyz} to constrain the shadowing region, generally resulting in a significantly stronger suppression of nuclear gluons at low $x_\mathrm{A}$~\cite{Eskola:2021nhw}. Notably, the di-hadron measurements are defined as the ratio of a two-hadron or two-jet (coincidence) cross-section to an inclusive hadron+X or jet+X cross-section.
The set of events with a trigger hadron or jet (selected inclusively, i.e., irrespective of the presence or properties of any additional hadrons or jets) generally arise from different $(x_\mathrm{A}, Q^2)$ distributions than the events with coincident pairs. Therefore, the suppression of these cross-sections from nPDF effects are not identical and may not fully cancel in the construction of these ratio observables. Given the recent updates to the nPDF sets, it is timely to investigate to what extent such di-hadron suppression signatures, commonly attributed to non-linear QCD saturation, could plausibly be described within a picture based on collinearly-factorized pQCD with an overall suppression of cross-sections based on nPDFs.

To explore this question concretely, I consider two recent measurements of forward di-hadron or di-jet correlations in $p$+A collisions, one from each of RHIC and the LHC: a measurement by ATLAS of forward di-jet conditional yields and azimuthal correlations~\cite{ATLAS:2019jgo} and a measurement by STAR of forward di-hadron correlation functions~\cite{STAR:2021fgw}. 
The focus on $p$+A rather than older $d$+A measurements avoids any potential complications associated with the deuteron probe~\cite{Strikman:2010bg}.
Both measurements report a significant suppression of the total per-trigger yield but, inconsistently with a saturation picture, an unmodified width of the azimuthal correlation function. The STAR measurement further compares two nuclear species, finding that the suppression is progressive in the nuclear size $A^{1/3}$. These two measurements do not attempt to select ``central'' events via activity in the $p$+A collision, and thus avoid any potential issues with multiplicity-selection biases~\cite{PHENIX:2013jxf,ALICE:2014xsp,Perepelitsa:2024eik}. The kinematic selections in the ATLAS measurement correspond to high $Q^2$ values where saturation effects were expected to be small (however, see Ref.~\cite{vanHameren:2019ysa}). Nevertheless, since it has qualitatively similar features as the STAR data, it is useful to consider the two together and evaluate to what extent a single framework could capture the features of both measurements.

The study presented here uses Monte Carlo (MC) event generators, matched to the specific kinematics of the measurements, with event-level reweighting according to the modifications provided by a selection of modern nPDF sets. As such, it is intended to serve as an initial assessment of how comparable the magnitude of expected nPDF effects may be with the signals observed in recent data. 
Implicitly, this is performed under the working assumption that the nPDF picture is appropriate at the low-$Q^2$ values probed by the measurements, even though higher-twist effects are expected to become increasingly important in this region and eventually violate these assumptions.
Additional work based on state-of-the-art theoretical calculations will be needed to critically examine the validity of these assumptions and thus the potential description of such observables within the different approaches to $p$+A (or $e$+A) collisions.

\section{Forward di-jet correlations at the LHC}
\label{sec:LHC}

The ATLAS measurement of di-jet correlations in Ref.~\cite{ATLAS:2019jgo} provides a good illustration of how nPDF effects may manifest, since there is a strong connection between the final-state jet kinematics and the parton-level kinematics, and the final states are produced by very high-$Q^2$ ($\gtrsim 100\  \mathrm{GeV}^2$) processes. 

ATLAS analyzes forward jet production in $\sqrt{s_\mathrm{NN}} = 5.02$~TeV $p$+Pb collisions and compares the result to that in same-energy $p$+$p$ collisions. Specifically, the measurement selects events with a leading (highest-$p_\mathrm{T}$, denoted $p_\mathrm{T,1}$) jet in the region $2.7 < y_1 < 4.0$, where positive values of the center-of-mass rapidity $y$ denote the proton-going direction, in different ranges of $p_\mathrm{T,1}$. Then, the per-trigger or conditional yield is determined for events which additionally have a sub-leading (next-highest-$p_\mathrm{T}$) jet as a function of $p_\mathrm{T,2}$ and $y_2$.

The per-trigger yield $I_{12}$ is defined as the number of coincident jet pairs $N_{12}$, divided by the total number of ``trigger'' leading jets $N_1$, where $N_1$ is determined without any restriction on the properties or even presence of a sub-leading partner jet. For events where the sub-leading jet is also required to be in the forward direction, for example $2.7 < y_2 < 4.0$, the ATLAS data shows an approximate 15\% suppression of the per-trigger yields in $p$+Pb collisions compared to the $p$+$p$ reference, $I_{12}^{p+\mathrm{Pb}} / I_{12}^{p+p}$. In addition, the difference in azimuthal angle between the jets, $\Delta\phi_{12}$, is reported and characterized with a root-mean square value $W_\mathrm{12}$. No modification of $W_\mathrm{12}$ between $p$+Pb and $p$+$p$ collisions is observed within uncertainties.

As suggested above, the definition of the ATLAS $I_{12}$ observable is potentially sensitive to a different magnitude of nPDF modification for inclusive trigger jet events ($N_1$) than for di-jet coincidence events ($N_{12}$), leading to a modification of the per-trigger yield but with minimal modification of the azimuthal correlation width $W_{12}$.

To evaluate the description of this measurement within an nPDF picture, the \textsc{Pythia} MC event generator~\cite{Bierlich:2022pfr}, version 8.307, was used to simulate the proton-nucleon collisions in $p$+Pb collisions which produce forward jets matching the ATLAS kinematic selections, and determine the distribution of $(x_\mathrm{A},Q^2)$, values for these events. 
This simulation was validated by comparing the per-trigger yields $I_{12}$ and the full $\Delta\phi$ distributions reported for $p$+$p$ collisions in Ref.~\cite{ATLAS:2019jgo} for different $p_{T,1}$, $p_{T,2}$, and $y_2$ selections. The \textsc{Pythia} simulation described these measured distributions in reasonable, but not perfect, detail over many orders of magnitude. This is because, although it includes many aspects which are important for a full description of hard processes such as initial and final state radiation and a sophisticated parton shower, it is not a true next-to-leading order generator.

The nuclear PDF effects were evaluated by determining, for each event, the modification factor $R_{f}^A(x_\mathrm{A},Q^2)$ for the flavor $f$ and Bjorken-$x$ of the hard-scattered parton in the nucleus $A$, $x_\mathrm{A}$, at the $Q^2$ of the \textsc{Pythia} hard process~\cite{Bierlich:2022pfr}. The modification factor was applied as an event-level weight. Thus, the impact of the nPDFs (such as on the trigger-jet or di-jet yields) in $p$+Pb collisions, compared to the $p$+$p$ collision baseline, can be evaluated by comparing the simulated events with and without this weighting.

Three recent nPDF sets for the $^{208}$Pb nucleus were considered: EPPS21~\cite{Eskola:2021nhw} as the principal one, with nCTEQ15~\cite{Kusina:2020lyz,Duwentaster:2021ioo} and nNNPDF 3.0~\cite{AbdulKhalek:2022fyi} as additional comparisons, each through the use of the LHAPDF~6.5.4 interface~\cite{Buckley:2014ana}. For EPPS21, LHAPDF provides the full PDF consisting of EPPS21 modifications atop the CT18A next-to-leading order (NLO) PDF~\cite{Hou:2019efy}, as well as the associated CT18ANLO free proton PDF. Thus, the modification factor was calculated by taking the ratio of the PDF values between the two. The uncertainty for EPPS21 was evaluated by considering the upper and lower limits of the 24 variations corresponding to the 90\% confidence level nuclear error set, and summing their impact in quadrature. The variations associated with the baseline CT18 errors are assumed to cancel strongly in the $p$+Pb-to-$p$+$p$ comparison and thus were not explicitly evaluated. For nCTEQ15, the ``WZ+SIH\_FullNuc'' version, was used. This version incorporates recent LHC $W$/$Z$ boson data and single inclusive hadron data, both to better describe the low-$x_\mathrm{A}$ gluon region, and which also accounts for the isospin composition of nucleons in the nucleus. For nNNPDF 3.0, the NLO version of the set with $\alpha_s = 0.118$ was used, as provided through LHAPDF.

\begin{figure}[!t]
\includegraphics[width=1.0\linewidth]{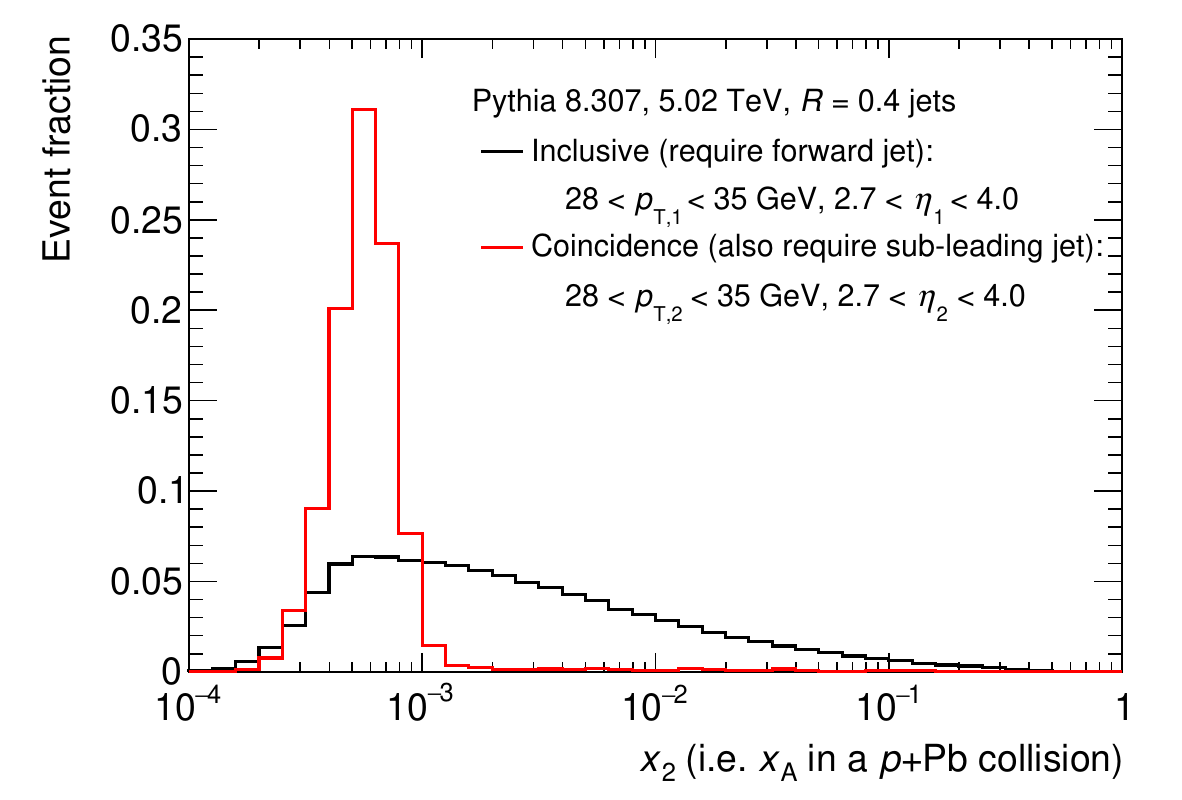}
\caption{\label{fig:xA_LHC} Distribution in \textsc{Pythia}8 of the values of Bjorken-$x$ in the nucleus in 5.02~TeV $p$+Pb collisions, for the ``inclusive'' selection of a leading jet with $p_\mathrm{T,1} = 28$--$35$~GeV in $2.7 < y_1 < 4.0$ and no other explicit requirement (black) and the ``coincidence'' selection where events must additionally contain a sub-leading jet with $p_\mathrm{T,2} = 28$--$35$~GeV in $2.7 < y_2 < 4.0$ (red).}
\end{figure}

To illustrate how nPDF effects manifest in the ATLAS measurement, Fig.~\ref{fig:xA_LHC} shows an example of $x_\mathrm{A}$ distributions for two selections: the forward trigger jet selection alone (the inclusive selection), and the addition of a specific forward sub-leading jet requirement (the coincidence selection). While the trigger jet selection predominantly selects low-$x_\mathrm{A}$ events, the distribution of $x_\mathrm{A}$ values is nevertheless broad. On the other hand, by constraining the kinematics of the full di-jet system with the coincidence selection, the resulting events have a sharply peaked $x_\mathrm{A}$ distribution around $x_\mathrm{A} \sim 5\times10^{-4}$. For the example kinematics here, the suppression in the total trigger jet yield $N_1$ from the nominal EPPS21 set is $\approx0.89$, while for the di-jet yield $N_{12}$ it is $\approx0.84$. Thus, the suppression of cross-sections from nPDF effects would partially but not completely cancel, and the per-trigger yield would be suppressed at a nominal level of $N_{12}/N_{1} \sim 0.84/0.89 \sim 0.94$. 

\begin{figure}[!t]
\includegraphics[width=1.0\linewidth]{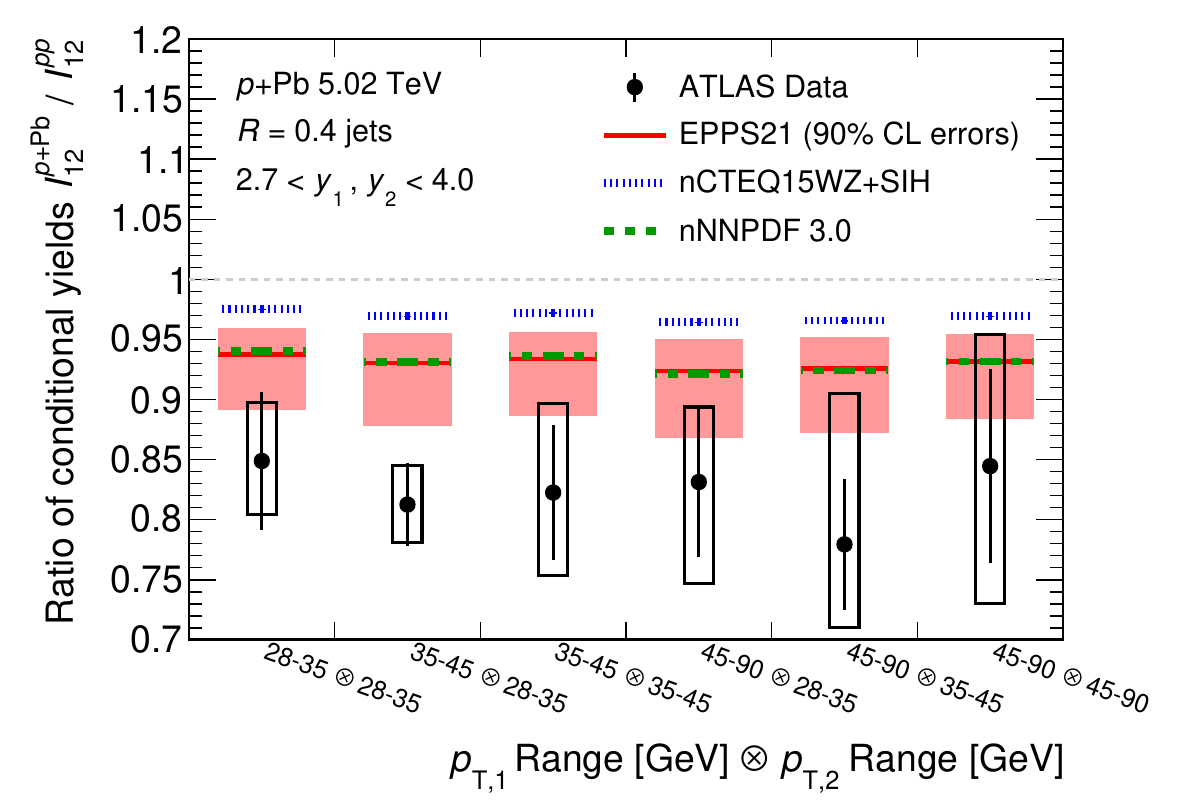}
\caption{\label{fig:LHC_Results1} Ratio of the per-trigger forward di-jet yield in $p$+Pb collisions to that in $p$+$p$ collisions at 5.02 TeV. The black points are data from ATLAS~\cite{ATLAS:2019jgo}, with the vertical lines and boxes showing the statistical and systematic uncertainties, respectively. The colored horizontal lines show respectively the values in EPPS21 (solid red), nCTEQ15WZ+SIH (dotted blue), and nNNPDF 3.0 (dashed green), with the shaded boxes around the EPPS21 line indicating the theoretical uncertainty (see text). Different selections on the leading and sub-leading jet $p_\mathrm{T}$ are shown at different horizontal positions, as indicated in the $x$-axis.}
\end{figure}

Fig.~\ref{fig:LHC_Results1} compares the ratio of the forward di-jet per-trigger yield $I_{12}$ in $p$+Pb to $p$+$p$ collisions, between the values measured in ATLAS to those calculated using \textsc{Pythia} plus EPPS21, nCTEQ15, and nNNPDF 3.0. The data compared here are for both jets in the forward-most selection $2.7 < y_1, y_2 < 4.0$, with the full set of $p_\mathrm{T,1}$ and $p_\mathrm{T,2}$ selections reported by ATLAS. The nominal suppression effect in  $I_{12}^{p\mathrm{+Pb}}/I_{12}^{p+p}$ from EPPS21 ($\approx7\%$) is approximately half of that in the nominal values of the data ($\approx15\%$). However, when considering the full statistical and systematic uncertainty in the data, as well as the uncertainty sets in EPPS21, in all but one kinematic bin, the data are fully compatible with the effect in this nPDF set. Ref.~\cite{ATLAS:2019jgo} notes that the experimental uncertainties in the $I_{12}$ ratio are dominated by those related to the jet energy scale, and its potential difference between $p$+Pb and $p$+$p$ datasets. Thus, while there is no explicit experimental guidance given on the point-to-point correlation, this uncertainty source could plausibly be strongly correlated between all the reported $p$+Pb/$p$+$p$ ratios.

As additional comparisons, the nominal nPDF effects in the nCTEQ15 and nNNPDF 3.0 sets have also been evaluated. The central value of nNNPDF 3.0 agrees closely with that from EPPS21. However, nCTEQ15 shows a much smaller level of suppression. This could arise from, for example, a different choice of input data  providing constraints on the low-$x_\mathrm{A}$ region of nPDF modification.

\begin{figure}[!t]
\includegraphics[width=1.0\linewidth]{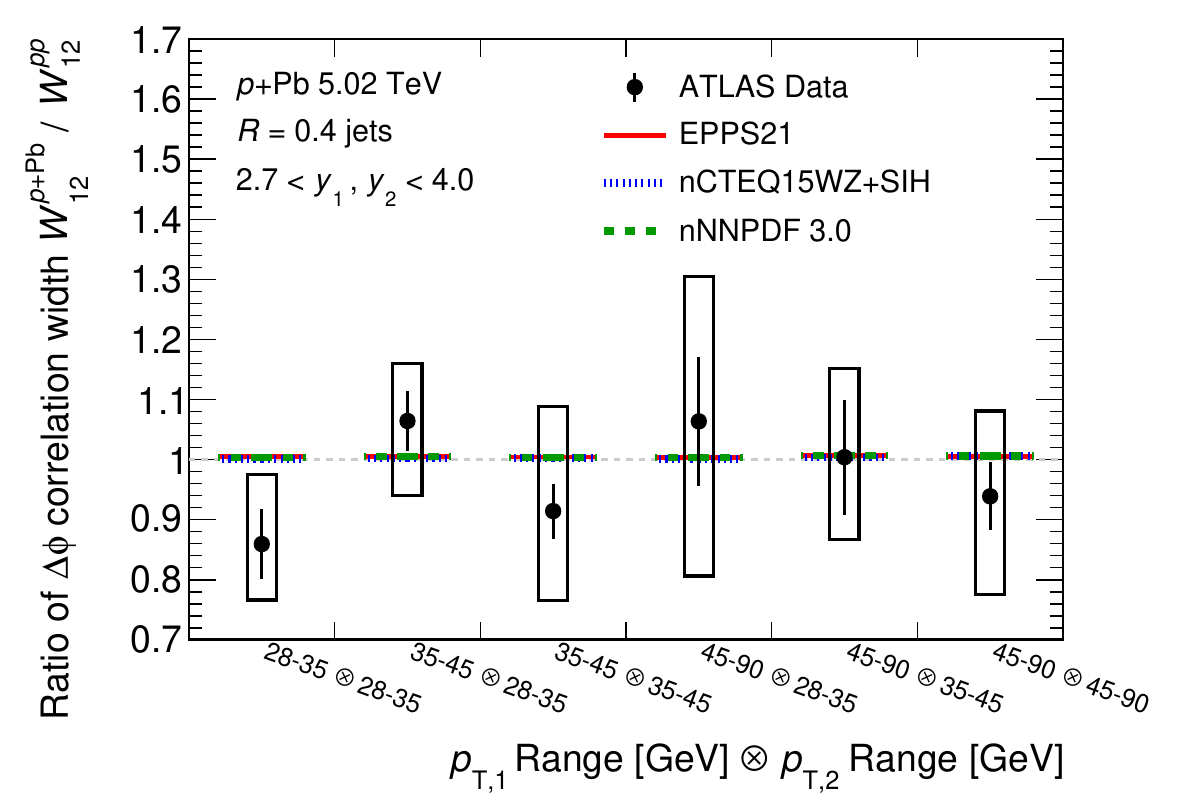}
\caption{\label{fig:LHC_Results2} Ratio of the forward di-jet azimuthal correlation width in $p$+Pb collisions to that in $p$+$p$ collisions at 5.02 TeV. The black points are data from ATLAS~\cite{ATLAS:2019jgo}, with the vertical lines and boxes showing the statistical and systematic uncertainties, respectively. The colored horizontal lines show respectively the values in EPPS21 (solid red), nCTEQ15WZ+SIH (dotted blue), and nNNPDF 3.0 (dashed green). nPDF uncertainties are not drawn. Different selections on the leading and sub-leading jet $p_\mathrm{T}$ are shown at different horizontal positions, as indicated in the $x$-axis.}
\end{figure}

For the same set of kinematic selections, Fig.~\ref{fig:LHC_Results2} compares the ratio of the forward di-jet correlation width $W_{12}$ in $p$+Pb to $p$+$p$ collisions between the measured values and those calculated using the EPPS21, nCTEQ15, and nNNPDF 3.0 nPDF sets. The nominal value of $W_{12}^{p\mathrm{+Pb}}/W_{12}^{p+p}$ for all sets is very nearly unity, well within the uncertainties of the data. The lack of any significant broadening may be expected because, in the collinear pQCD plus nPDF picture, the properties of individual events themselves are unmodified, but rather the events contribute to the observable with a modified overall weight.

Thus, an nPDF-based picture, based purely on reweighting cross-sections according to their $(x_\mathrm{A},Q^2)$ values, is able to describe the small but significant suppression in the per-trigger yield observed in ATLAS data while also naturally resulting in no modification to the width of the $\Delta\phi$ distribution, and it is able to do so consistently for different selections on the kinematics of the two jets.

\section{Forward di-hadron correlations at RHIC}

The STAR measurement of forward di-hadron correlations in $p$+Au collisions~\cite{STAR:2021fgw} is more challenging to model, due to the weaker connection between measured hadron fragments and the initial-state kinematics, the lower $Q^2$ values which are close to the limits of validity for global nPDF extractions, and other issues such as the presence of an uncorrelated di-hadron pedestal. Thus, while the modeling is broadly similar to that for di-jets described in Sec.~\ref{sec:LHC}, some key differences are highlighted below.

STAR measures forward di-hadron production in $\sqrt{s_\mathrm{NN}} = 200$~GeV $p$+Au and $p$+Al collisions and compares the result to that in same-energy $p$+$p$ collisions. The measurement selects ``trigger'' $\pi^0$ hadrons in the forward region $2.6 < \eta < 4.0$, for multiple $p_\mathrm{T}$ ranges denoted $p_\mathrm{T}^\mathrm{trig}$, with a total yield $N_1$. The pair yield $N_{12}$ for additionally finding an ``associated'' $\pi^0$ in the same pseudorapidity range, but in a lower $p_\mathrm{T}$ range ($p_\mathrm{T}^\mathrm{assoc}$), is measured. The correlation function is defined as the per-trigger yield $N_{12} / N_{1}$ and is reported as a function of $\Delta\phi$ between the $\pi^0$ pair. Thus again the measured observable is the ratio of a coincidence cross-section to an inclusive cross-section. Unlike the case with di-jets, where the $\Delta\phi$ function is completely dominated by a strong back-to-back peak, the di-hadron correlation includes a significant pedestal contribution flat in $\Delta\phi$ which could arise from, for example, underlying event (UE) production, multi-parton interactions (MPIs), or double parton scatterings~\cite{Strikman:2010bg}. The pair yield is contained in the back-to-back contribution peaked at $\Delta\phi = \pi$, which must be separated from this pedestal, thus introducing some level of uncertainty in the measurement and also in the modeling applied here. 

STAR reports the area and width of the back-to-back component of the correlation function in $p$+Au and $p$+Al collisions compared to that in $p$+$p$, under different selections in $p_\mathrm{T}^\mathrm{trig}$ and $p_\mathrm{T}^\mathrm{assoc}$. The area (i.e. yield) shows a strong suppression of 20--40\% depending on the collision system and particular kinematic selections. On the other hand, the width of the correlation function is unmodified within uncertainties between the $p$+A systems and $p$+$p$.

As discussed above, nPDF effects on the total trigger yield ($N_1$) will generally be different than that on the pair yield ($N_{12}$), since these sample different distributions of $x_\mathrm{A}$ and $Q^2$. Additionally, the modification may be $\Delta\phi$-dependent, since different types of processes  contribute in different $\Delta\phi$ regions. 

For the modeling of this data, the simulation setup was replicated from that specifically used by STAR in their measurement~\cite{STAR:2021fgw,Private}.
The \textsc{Pythia} generator version 6.428~\cite{Sjostrand:2006za} was configured with Tune 370 (Perugia 2012)~\cite{Skands:2010ak} and the CTEQ6L1~\cite{Pumplin:2002vw} PDF set, and with low-$p_\mathrm{T}$ production and semi-hard QCD $2\to2$ processes enabled.
Notably, the LO PDF in this setup is used with nPDF sets extracted with respect to NLO PDFs. While a check of the impact of LO vs. NLO PDF sets in \textsc{Pythia} suggested that this does not have a major impact on the essential result, it highlights the need for a true NLO calculation in future studies. 
The simulated events result in correlation functions which are similar to those measured in $p$+$p$ data, with a slightly smaller correlation width (by 10--20\% depending on the kinematic selection) but a significantly lower pedestal.
For this study, the correlated yield is extracted through a Gaussian plus constant pedestal fit to the correlation function. These fits give good descriptions of the distributions, with some example fits shown below.

The nPDF effects on the measurement were calculated in a similar way as in Sec.~\ref{sec:LHC}, but using the analogous sets for the $^{197}$Au and $^{27}$Al nuclei. For the EPPS21 set specifically, there is a rare technical problem, also noted in Ref.~\cite{Eskola:2021nhw}, in which some nPDFs can evaluate to negative values at very low $Q^2$ values. Thus, following the suggestion of the EPPS21 authors, for this nPDF set in particular the modifications are determined using either the $Q^2$ which is the hard-process $Q^2$ given by \textsc{Pythia} or $1.8$~GeV$^2$, whichever is greater. 

\begin{figure}[!t]
\includegraphics[width=1.0\linewidth]{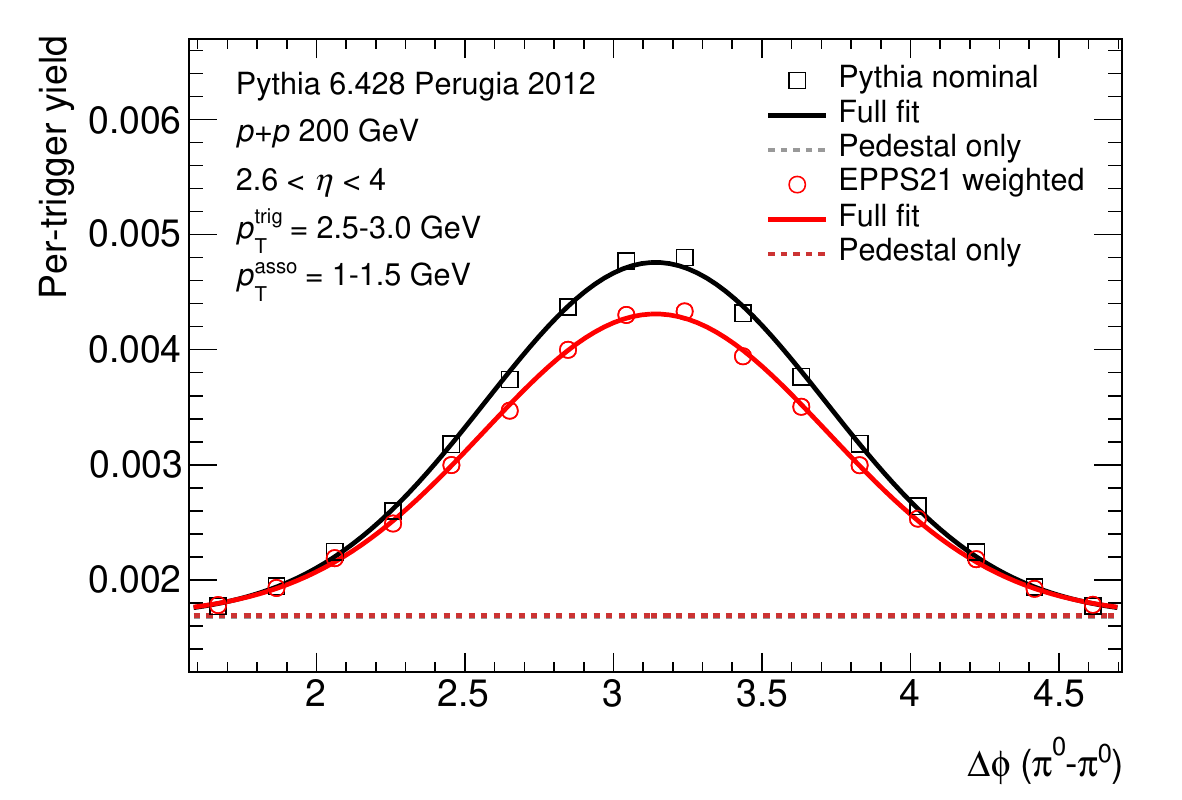}
\caption{\label{fig:dphi_RHIC} Per-trigger normalized correlation function for forward di-hadrons, with an example kinematic selection $p_\mathrm{T}^\mathrm{trig} = 2$--$2.5$~GeV and $p_\mathrm{T}^\mathrm{assoc} = 1$--$1.5$~GeV. The open black square and open red circle markers show the distribution in nominal \textsc{Pythia} events compared to that after event-level reweighting by the EPPS21 nPDF set, respectively (see text). The associated solid lines show the results of a Gaussian plus constant pedestal fit to the distribution, while the dashed lines show just the pedestal component.
}
\end{figure}

To illustrate how nPDF effects manifest in the STAR measurement, Fig.~\ref{fig:dphi_RHIC} shows an example correlation function matched to one of the STAR kinematic selections, with and without EPPS21 reweighting. In a similar vein as the forward di-jet situation described in Sec.~\ref{sec:LHC}, events with a selected trigger hadron sample a broad distribution of $x_\mathrm{A}$ values extending to low values, and thus the total trigger yield $N_1$ is significantly suppressed by nPDF effects. However, the events which additionally have an associated back-to-back hadron have a different $x_\mathrm{A}$ distribution and the correlated pair yield $N_{12}$ exhibits an even stronger nPDF suppression. Thus, these do not cancel in the per-trigger yield and nPDF effects result in a suppression of this observable as well, as seen in Fig.~\ref{fig:dphi_RHIC}. Interestingly, the pedestal in the per-trigger yield is essentially unaffected when applying an nPDF weighting according to the hard process kinematics, which makes sense if it arises from MPIs or the UE more generally. In the STAR data, the extracted pedestal in $p$+Au and $p$+Al data is similarly consistent with that in $p$+$p$ data, with the exception of a small decrease in the lowest $p_\mathrm{T}^\mathrm{assoc}$ selection.

\begin{figure}[!t]
\includegraphics[width=1.0\linewidth]{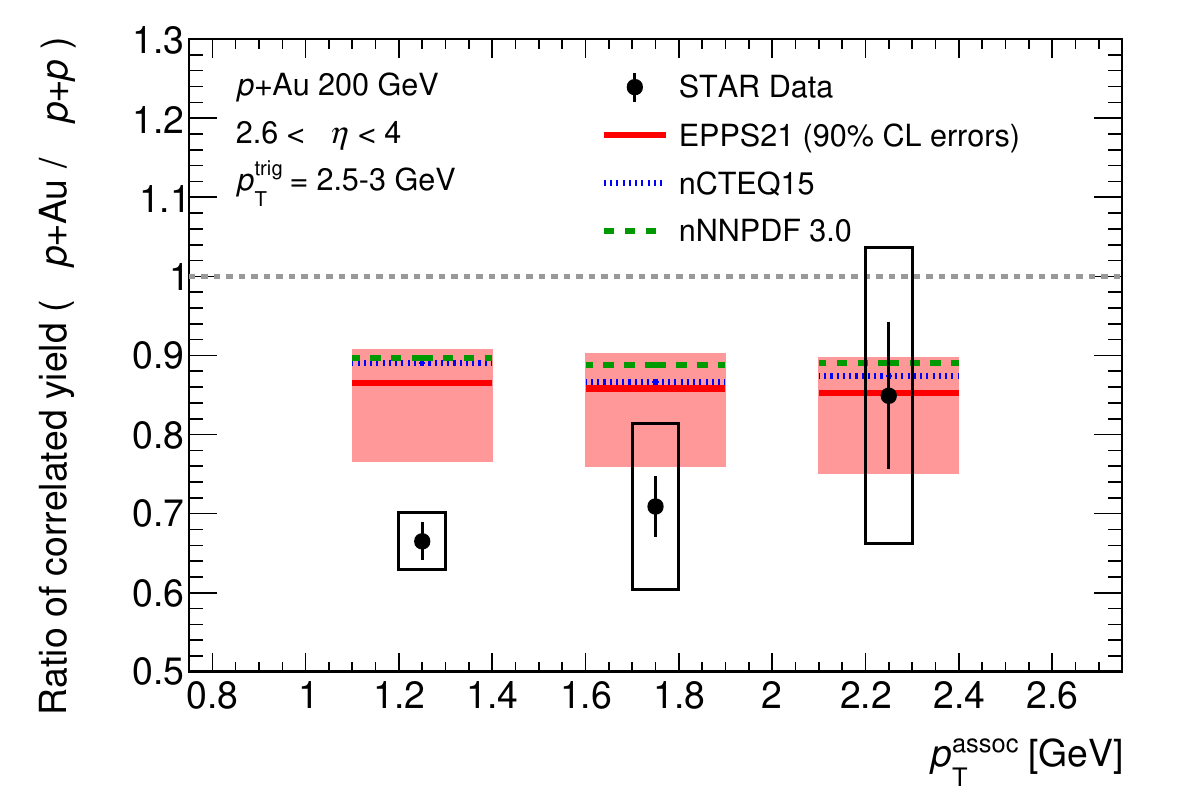}
\caption{\label{fig:RHIC_Results1} Ratio of the forward dihadron azimuthally-correlated yield in $p$+Au collisions to that in $p$+$p$ collisions at 200 GeV, as a function of associated hadron $p_\mathrm{T}^\mathrm{assoc}$. The black points are data from STAR~\cite{STAR:2021fgw}, with the vertical lines and boxes showing the statistical and systematic uncertainties, respectively. The colored horizontal lines show respectively the values in EPPS21 (solid red), nCTEQ15WZ+SIH (dotted blue), and nNNPDF 3.0 (dashed green), with the shaded boxes around the EPPS21 line indicating the theoretical uncertainty (see text).}
\end{figure}

Fig.~\ref{fig:RHIC_Results1} compares the ratio of the forward di-hadron per-trigger correlated yield in $p$+Au to $p$+$p$ collisions, between the values measured in STAR to those calculated using \textsc{Pythia} plus EPPS21, nCTEQ15, and nNNPDF 3.0. The data here are for trigger hadrons with $p_\mathrm{T} = 2$--$2.5$~GeV, corresponding to Fig.~2 of Ref.~\cite{STAR:2021fgw}. In the lowest $p_\mathrm{T}^\mathrm{assoc}$ selection (1--1.5~GeV), the nominal value of this observable within EPPS21 has approximately half of the suppression effect in the nominal value of the data, although it could be compatible with the majority of the effect within the combined experimental and theoretical uncertainties in the data and the nPDF set. For the other two $p_\mathrm{T}^\mathrm{assoc}$ selections, the suppression of the conditional yield is fully compatible with the nPDF-based calculation. For the two additional nPDF sets considered here, nCTEQ15 and nNNPDF 3.0, the nominal suppression effects are present but somewhat smaller than that in EPPS21. I note that additional uncertainties in the modeling which have not been evaluated, such as those related to the correlated yield extraction or the imperfect data-simulation agreement in the correlation function, may make the nPDF calculations further compatible with the data.

\begin{figure}[!t]
\includegraphics[width=1.0\linewidth]{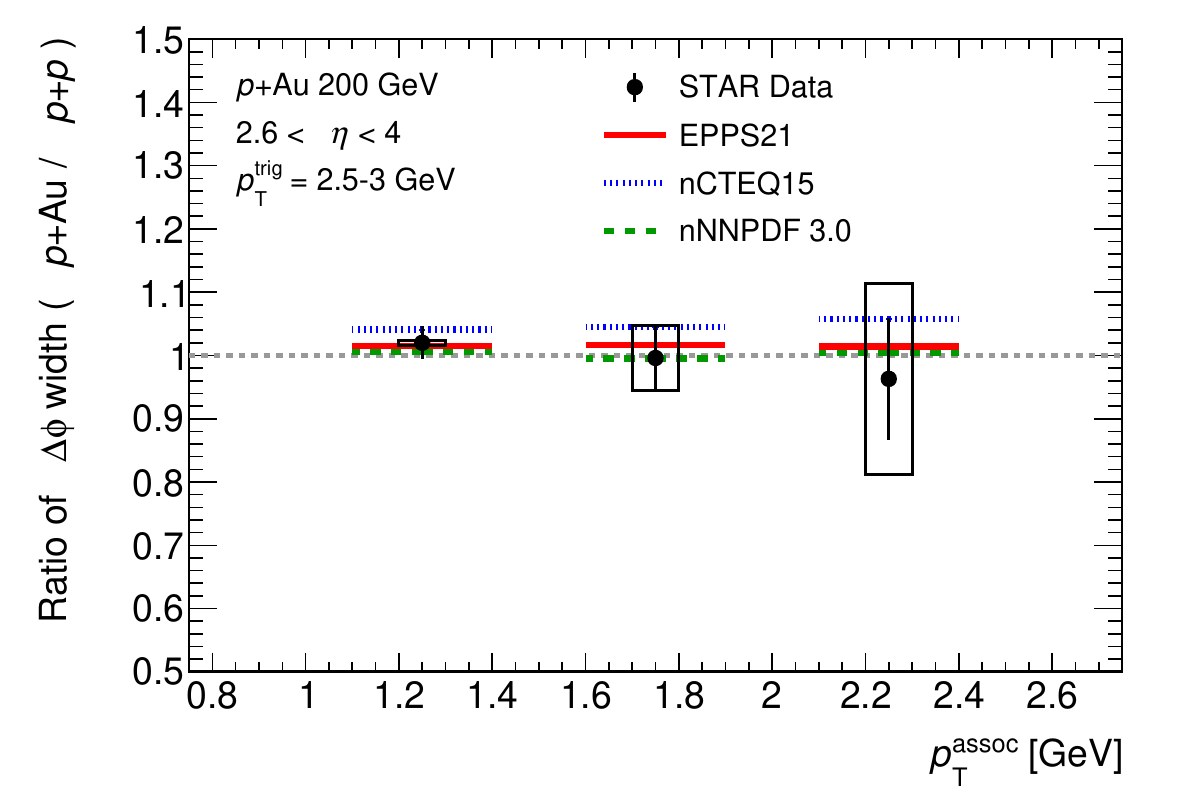}
\caption{\label{fig:RHIC_Results2} Ratio of the forward dihadron azimuthal correlation width in $p$+Au collisions to that in $p$+$p$ collisions at 200 GeV, as a function of associated hadron $p_\mathrm{T}$. The black points are data from STAR~\cite{STAR:2021fgw}, with the vertical lines and boxes showing the statistical and systematic uncertainties, respectively. The colored horizontal lines show respectively the values in EPPS21 (solid red), nCTEQ15WZ+SIH (dotted blue), and nNNPDF 3.0 (dashed green). nPDF uncertainties are not drawn. }
\end{figure}

For the same set of kinematic selections, Fig.~\ref{fig:RHIC_Results2} compares the ratio of the forward di-hadron correlation width in $p$+Au to $p$+$p$ collisions between the measured values and those calculated using EPPS21, nCTEQ15, and nNNPDF 3.0. As also seen in the LHC di-jet case above, the nominal value of this ratio for all sets is very nearly unity, and again well within the uncertainties of the data. Interestingly, the nominal effect within nCTEQ15 is to have a modest (5\%) increase in the correlation width, which may arise from the relative reweighting of event classes which themselves have different correlation widths. This suggests that even small broadening effects (which have not yet been observed in the data) could be potentially accommodated within an nPDF picture as well.

\begin{figure}[!t]
\includegraphics[width=1.0\linewidth]{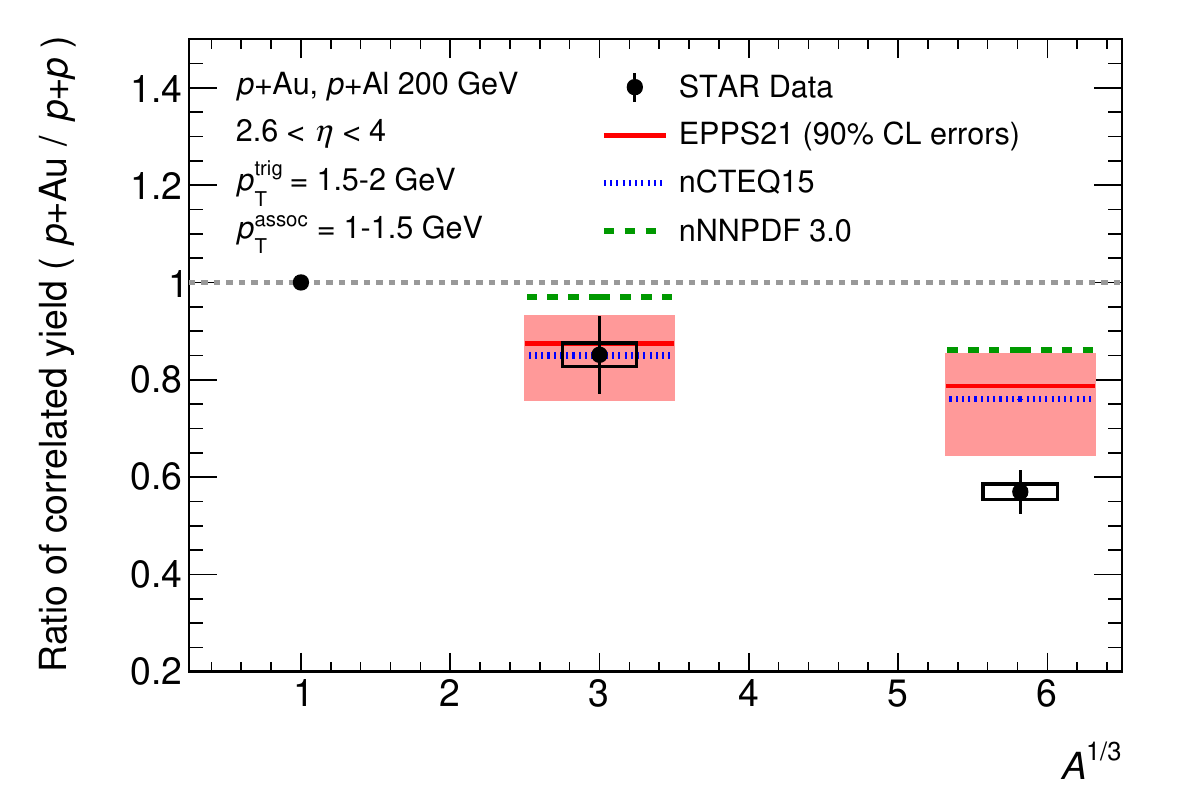}
\caption{\label{fig:RHIC_Results3} Ratio of the forward dihadron azimuthally-correlated yield in $p$+A collisions to that in $p$+$p$ collisions at 200 GeV, as a function of nuclear size $A^{1/3}$. The black points are data from STAR~\cite{STAR:2021fgw}, with the vertical lines and boxes showing the statistical and systematic uncertainties, respectively. The colored horizontal lines show respectively the values in EPPS21 (solid red), nCTEQ15WZ+SIH (dotted blue), and nNNPDF 3.0 (dashed green), with the boxes around the EPPS21 line indicating the theoretical uncertainty (see text).}
\end{figure}

Fig.~\ref{fig:RHIC_Results3} compares the $A$ dependence of the ratio of the forward di-hadron per-trigger correlated yield in $p$+A to $p$+$p$ collisions. For this kinematic selection, the nPDF effects as implemented in EPPS21 fully describe the suppression in $p$+Al data and are compatible with the large majority of the effect in $p$+Au data. Thus, the magnitude of the suppression in data follows a similar quantitative trend as might be expected from the $A$-dependent shadowing of cross-sections, at least as incorporated in global nPDF extractions. The nCTEQ15 set gives similar overall values as EPPS21, while nNNPDF 3.0 generates a smaller suppression for both collision systems.

The systematic data comparisons above suggest that a straightforward nPDF-based picture, arising from the differential suppression of cross-sections, can potentially describe a significant fraction of the suppression in the conditional yield observed in STAR data, or even the full effect, depending on the particular kinematic selections. This mechanism also naturally explains the lack of significant modification in the width of the $\Delta\phi$ distribution, and it is able to describe these features in the data for different selections on the hadron kinematics and even collision species.

\section{Conclusion}

Di-hadron and di-jet correlation measurements in $p$+A collisions are consistently motivated as revealing probes of novel, non-linear QCD dynamics inside dense nuclei.
Somewhat unexpectedly, pictures based on leading-twist perturbative QCD within a collinear factorization framework, and using a universal set of $(x_\mathrm{A},Q^2)$-dependent modifications of parton densities in nuclei, seem to be able to describe major features of recent forward di-hadron and di-jet data.
This is striking because the nPDF picture has no mechanism for changing the nature of inter-event correlations beyond a simple $(x_\mathrm{A},Q^2)$-dependent reweighting of overall cross-sections. 
Interestingly, the nPDF paradigm also gives a natural explanation for why conditional yields are suppressed without a significant change in the correlation width, as would qualitatively be expected from the saturation picture. The remaining difference between the suppression effect captured by nPDFs and the suppression in the data could be explained by additional dynamical QCD effects, but constraining the magnitude of these would require significant improvements in the experimental and theoretical (nPDF) uncertainties. Thus di-hadron and di-jet observables alone, if defined in this way, may not provide a robust way to identify saturation phenomena. 

This finding highlights a broader question in the heavy-ion physics community, which is to what extent different paradigms describing hard processes in $p$+A-style collisions may be expected to overlap. For example, first-principles calculations of saturation phenomena may result in suppressed gluon densities at low-$x_\mathrm{A}$, i.e. consistent with the experimental phenomenon of shadowing of cross-sections. Going in the other direction, global nPDF analyses are agnostic as to the specific physical origin of nuclear effects, and faithfully encode them into the extracted nPDF sets. Thus the two approaches do not, in principle, capture mutually exclusive physics~\cite{Boussarie:2016bkq,Boussarie:2021ybe,Armesto:2022mxy,Schenke:2024gnj,Cheung:2024qvw}. For the isolation of non-linear QCD saturation effects, it may thus be important to identify the specific saturation scale $Q_s$ at which the assumed universality, or other assumptions, of the nPDF picture are observed to break down. Furthermore, it highlights the necessity of multiple, corroborating observables to definitively observe exotic gluon saturation behavior in high-energy nuclear collisions.

Finally, this study performed an initial assessment of two particular recent measurements with a simple Monte Carlo-based approach. In the future, it would be interesting to approach this problem with a more sophisticated theoretical treatment, including the evaluation of higher-order effects. It may also be interesting to study other di-hadron/jet or photon-hadron/jet~\cite{Benic:2022ixp} correlation measurements within the context of the latest nPDF sets, and evaluate what these may predict for measurements in $e$+A collisions at the EIC, where the probed $(x_\mathrm{A},Q^2)$ values can be experimentally selected.

\begin{acknowledgments}
DVP acknowledges Xiaoxuan Chu, Peter Jacobs, Jamal Jalilian-Marian, John Lajoie, Riccardo Longo, Farid Salazar, Bjoern Schenke, and Ivan Vitev for useful discussions and input on an earlier draft of this paper. 
DVP further thanks the Institute for Nuclear Theory at the University of Washington for its kind hospitality and stimulating research environment, particularly the organizers and attendees of INT Program 24-2b, Heavy Ion Physics in the EIC Era, where the idea behind this study was initially developed. 
DVP's work is supported by the U.S. Department of Energy, grant No. DE-FG02-03ER41244, and in part by the INT's U.S. DOE grant No. DE-FG02-00ER41132.
\end{acknowledgments}

\bibliography{apssamp}

\begin{thebibliography}{53}%
\makeatletter
\providecommand \@ifxundefined [1]{%
 \@ifx{#1\undefined}
}%
\providecommand \@ifnum [1]{%
 \ifnum #1\expandafter \@firstoftwo
 \else \expandafter \@secondoftwo
 \fi
}%
\providecommand \@ifx [1]{%
 \ifx #1\expandafter \@firstoftwo
 \else \expandafter \@secondoftwo
 \fi
}%
\providecommand \natexlab [1]{#1}%
\providecommand \enquote  [1]{``#1''}%
\providecommand \bibnamefont  [1]{#1}%
\providecommand \bibfnamefont [1]{#1}%
\providecommand \citenamefont [1]{#1}%
\providecommand \href@noop [0]{\@secondoftwo}%
\providecommand \href [0]{\begingroup \@sanitize@url \@href}%
\providecommand \@href[1]{\@@startlink{#1}\@@href}%
\providecommand \@@href[1]{\endgroup#1\@@endlink}%
\providecommand \@sanitize@url [0]{\catcode `\\12\catcode `\$12\catcode
  `\&12\catcode `\#12\catcode `\^12\catcode `\_12\catcode `\%12\relax}%
\providecommand \@@startlink[1]{}%
\providecommand \@@endlink[0]{}%
\providecommand \url  [0]{\begingroup\@sanitize@url \@url }%
\providecommand \@url [1]{\endgroup\@href {#1}{\urlprefix }}%
\providecommand \urlprefix  [0]{URL }%
\providecommand \Eprint [0]{\href }%
\providecommand \doibase [0]{https://doi.org/}%
\providecommand \selectlanguage [0]{\@gobble}%
\providecommand \bibinfo  [0]{\@secondoftwo}%
\providecommand \bibfield  [0]{\@secondoftwo}%
\providecommand \translation [1]{[#1]}%
\providecommand \BibitemOpen [0]{}%
\providecommand \bibitemStop [0]{}%
\providecommand \bibitemNoStop [0]{.\EOS\space}%
\providecommand \EOS [0]{\spacefactor3000\relax}%
\providecommand \BibitemShut  [1]{\csname bibitem#1\endcsname}%
\let\auto@bib@innerbib\@empty
\bibitem [{\citenamefont {Salgado}\ \emph {et~al.}(2012)\citenamefont {Salgado}
  \emph {et~al.}}]{Salgado:2011wc}%
  \BibitemOpen
  \bibfield  {author} {\bibinfo {author} {\bibfnamefont {C.~A.}\ \bibnamefont
  {Salgado}} \emph {et~al.},\ }\bibfield  {title} {\bibinfo {title}
  {{Proton-Nucleus Collisions at the LHC: Scientific Opportunities and
  Requirements}},\ }\href {https://doi.org/10.1088/0954-3899/39/1/015010}
  {\bibfield  {journal} {\bibinfo  {journal} {J. Phys. G}\ }\textbf {\bibinfo
  {volume} {39}},\ \bibinfo {pages} {015010} (\bibinfo {year} {2012})},\
  \Eprint {https://arxiv.org/abs/1105.3919} {arXiv:1105.3919 [hep-ph]}
  \BibitemShut {NoStop}%
\bibitem [{\citenamefont {Salgado}\ and\ \citenamefont
  {Wessels}(2016)}]{Salgado:2016jws}%
  \BibitemOpen
  \bibfield  {author} {\bibinfo {author} {\bibfnamefont {C.~A.}\ \bibnamefont
  {Salgado}}\ and\ \bibinfo {author} {\bibfnamefont {J.~P.}\ \bibnamefont
  {Wessels}},\ }\bibfield  {title} {\bibinfo {title} {{Proton\textendash{}Lead
  Collisions at the CERN LHC}},\ }\href
  {https://doi.org/10.1146/annurev-nucl-102014-022110} {\bibfield  {journal}
  {\bibinfo  {journal} {Ann. Rev. Nucl. Part. Sci.}\ }\textbf {\bibinfo
  {volume} {66}},\ \bibinfo {pages} {449} (\bibinfo {year} {2016})}\BibitemShut
  {NoStop}%
\bibitem [{\citenamefont {Ethier}\ and\ \citenamefont
  {Nocera}(2020)}]{Ethier:2020way}%
  \BibitemOpen
  \bibfield  {author} {\bibinfo {author} {\bibfnamefont {J.~J.}\ \bibnamefont
  {Ethier}}\ and\ \bibinfo {author} {\bibfnamefont {E.~R.}\ \bibnamefont
  {Nocera}},\ }\bibfield  {title} {\bibinfo {title} {{Parton Distributions in
  Nucleons and Nuclei}},\ }\href
  {https://doi.org/10.1146/annurev-nucl-011720-042725} {\bibfield  {journal}
  {\bibinfo  {journal} {Ann. Rev. Nucl. Part. Sci.}\ }\textbf {\bibinfo
  {volume} {70}},\ \bibinfo {pages} {43} (\bibinfo {year} {2020})},\ \Eprint
  {https://arxiv.org/abs/2001.07722} {arXiv:2001.07722 [hep-ph]} \BibitemShut
  {NoStop}%
\bibitem [{\citenamefont {Qiu}\ and\ \citenamefont {Vitev}(2004)}]{Qiu:2003vd}%
  \BibitemOpen
  \bibfield  {author} {\bibinfo {author} {\bibfnamefont {J.-w.}\ \bibnamefont
  {Qiu}}\ and\ \bibinfo {author} {\bibfnamefont {I.}~\bibnamefont {Vitev}},\
  }\bibfield  {title} {\bibinfo {title} {{Resummed QCD power corrections to
  nuclear shadowing}},\ }\href {https://doi.org/10.1103/PhysRevLett.93.262301}
  {\bibfield  {journal} {\bibinfo  {journal} {Phys. Rev. Lett.}\ }\textbf
  {\bibinfo {volume} {93}},\ \bibinfo {pages} {262301} (\bibinfo {year}
  {2004})},\ \Eprint {https://arxiv.org/abs/hep-ph/0309094}
  {arXiv:hep-ph/0309094} \BibitemShut {NoStop}%
\bibitem [{\citenamefont {Qiu}\ and\ \citenamefont {Vitev}(2006)}]{Qiu:2004da}%
  \BibitemOpen
  \bibfield  {author} {\bibinfo {author} {\bibfnamefont {J.-w.}\ \bibnamefont
  {Qiu}}\ and\ \bibinfo {author} {\bibfnamefont {I.}~\bibnamefont {Vitev}},\
  }\bibfield  {title} {\bibinfo {title} {{Coherent QCD multiple scattering in
  proton-nucleus collisions}},\ }\href
  {https://doi.org/10.1016/j.physletb.2005.10.073} {\bibfield  {journal}
  {\bibinfo  {journal} {Phys. Lett. B}\ }\textbf {\bibinfo {volume} {632}},\
  \bibinfo {pages} {507} (\bibinfo {year} {2006})},\ \Eprint
  {https://arxiv.org/abs/hep-ph/0405068} {arXiv:hep-ph/0405068} \BibitemShut
  {NoStop}%
\bibitem [{\citenamefont {McLerran}\ and\ \citenamefont
  {Venugopalan}(1994{\natexlab{a}})}]{McLerran:1993ni}%
  \BibitemOpen
  \bibfield  {author} {\bibinfo {author} {\bibfnamefont {L.~D.}\ \bibnamefont
  {McLerran}}\ and\ \bibinfo {author} {\bibfnamefont {R.}~\bibnamefont
  {Venugopalan}},\ }\bibfield  {title} {\bibinfo {title} {{Computing quark and
  gluon distribution functions for very large nuclei}},\ }\href
  {https://doi.org/10.1103/PhysRevD.49.2233} {\bibfield  {journal} {\bibinfo
  {journal} {Phys. Rev. D}\ }\textbf {\bibinfo {volume} {49}},\ \bibinfo
  {pages} {2233} (\bibinfo {year} {1994}{\natexlab{a}})},\ \Eprint
  {https://arxiv.org/abs/hep-ph/9309289} {arXiv:hep-ph/9309289} \BibitemShut
  {NoStop}%
\bibitem [{\citenamefont {McLerran}\ and\ \citenamefont
  {Venugopalan}(1994{\natexlab{b}})}]{McLerran:1993ka}%
  \BibitemOpen
  \bibfield  {author} {\bibinfo {author} {\bibfnamefont {L.~D.}\ \bibnamefont
  {McLerran}}\ and\ \bibinfo {author} {\bibfnamefont {R.}~\bibnamefont
  {Venugopalan}},\ }\bibfield  {title} {\bibinfo {title} {{Gluon distribution
  functions for very large nuclei at small transverse momentum}},\ }\href
  {https://doi.org/10.1103/PhysRevD.49.3352} {\bibfield  {journal} {\bibinfo
  {journal} {Phys. Rev. D}\ }\textbf {\bibinfo {volume} {49}},\ \bibinfo
  {pages} {3352} (\bibinfo {year} {1994}{\natexlab{b}})},\ \Eprint
  {https://arxiv.org/abs/hep-ph/9311205} {arXiv:hep-ph/9311205} \BibitemShut
  {NoStop}%
\bibitem [{\citenamefont {Gelis}\ \emph {et~al.}(2010)\citenamefont {Gelis},
  \citenamefont {Iancu}, \citenamefont {Jalilian-Marian},\ and\ \citenamefont
  {Venugopalan}}]{Gelis:2010nm}%
  \BibitemOpen
  \bibfield  {author} {\bibinfo {author} {\bibfnamefont {F.}~\bibnamefont
  {Gelis}}, \bibinfo {author} {\bibfnamefont {E.}~\bibnamefont {Iancu}},
  \bibinfo {author} {\bibfnamefont {J.}~\bibnamefont {Jalilian-Marian}},\ and\
  \bibinfo {author} {\bibfnamefont {R.}~\bibnamefont {Venugopalan}},\
  }\bibfield  {title} {\bibinfo {title} {{The Color Glass Condensate}},\ }\href
  {https://doi.org/10.1146/annurev.nucl.010909.083629} {\bibfield  {journal}
  {\bibinfo  {journal} {Ann. Rev. Nucl. Part. Sci.}\ }\textbf {\bibinfo
  {volume} {60}},\ \bibinfo {pages} {463} (\bibinfo {year} {2010})},\ \Eprint
  {https://arxiv.org/abs/1002.0333} {arXiv:1002.0333 [hep-ph]} \BibitemShut
  {NoStop}%
\bibitem [{\citenamefont {Morreale}\ and\ \citenamefont
  {Salazar}(2021)}]{Morreale:2021pnn}%
  \BibitemOpen
  \bibfield  {author} {\bibinfo {author} {\bibfnamefont {A.}~\bibnamefont
  {Morreale}}\ and\ \bibinfo {author} {\bibfnamefont {F.}~\bibnamefont
  {Salazar}},\ }\bibfield  {title} {\bibinfo {title} {{Mining for Gluon
  Saturation at Colliders}},\ }\href {https://doi.org/10.3390/universe7080312}
  {\bibfield  {journal} {\bibinfo  {journal} {Universe}\ }\textbf {\bibinfo
  {volume} {7}},\ \bibinfo {pages} {312} (\bibinfo {year} {2021})},\ \Eprint
  {https://arxiv.org/abs/2108.08254} {arXiv:2108.08254 [hep-ph]} \BibitemShut
  {NoStop}%
\bibitem [{\citenamefont {Kharzeev}\ \emph {et~al.}(2005)\citenamefont
  {Kharzeev}, \citenamefont {Levin},\ and\ \citenamefont
  {McLerran}}]{Kharzeev:2004bw}%
  \BibitemOpen
  \bibfield  {author} {\bibinfo {author} {\bibfnamefont {D.}~\bibnamefont
  {Kharzeev}}, \bibinfo {author} {\bibfnamefont {E.}~\bibnamefont {Levin}},\
  and\ \bibinfo {author} {\bibfnamefont {L.}~\bibnamefont {McLerran}},\
  }\bibfield  {title} {\bibinfo {title} {{Jet azimuthal correlations and parton
  saturation in the color glass condensate}},\ }\href
  {https://doi.org/10.1016/j.nuclphysa.2004.10.031} {\bibfield  {journal}
  {\bibinfo  {journal} {Nucl. Phys. A}\ }\textbf {\bibinfo {volume} {748}},\
  \bibinfo {pages} {627} (\bibinfo {year} {2005})},\ \Eprint
  {https://arxiv.org/abs/hep-ph/0403271} {arXiv:hep-ph/0403271} \BibitemShut
  {NoStop}%
\bibitem [{\citenamefont {Jalilian-Marian}\ and\ \citenamefont
  {Kovchegov}(2004)}]{Jalilian-Marian:2004vhw}%
  \BibitemOpen
  \bibfield  {author} {\bibinfo {author} {\bibfnamefont {J.}~\bibnamefont
  {Jalilian-Marian}}\ and\ \bibinfo {author} {\bibfnamefont {Y.~V.}\
  \bibnamefont {Kovchegov}},\ }\bibfield  {title} {\bibinfo {title} {{Inclusive
  two-gluon and valence quark-gluon production in DIS and pA}},\ }\href
  {https://doi.org/10.1103/PhysRevD.71.079901} {\bibfield  {journal} {\bibinfo
  {journal} {Phys. Rev. D}\ }\textbf {\bibinfo {volume} {70}},\ \bibinfo
  {pages} {114017} (\bibinfo {year} {2004})},\ \bibinfo {note} {[Erratum:
  Phys.Rev.D 71, 079901 (2005)]},\ \Eprint
  {https://arxiv.org/abs/hep-ph/0405266} {arXiv:hep-ph/0405266} \BibitemShut
  {NoStop}%
\bibitem [{\citenamefont {Marquet}(2007)}]{Marquet:2007vb}%
  \BibitemOpen
  \bibfield  {author} {\bibinfo {author} {\bibfnamefont {C.}~\bibnamefont
  {Marquet}},\ }\bibfield  {title} {\bibinfo {title} {{Forward inclusive dijet
  production and azimuthal correlations in p(A) collisions}},\ }\href
  {https://doi.org/10.1016/j.nuclphysa.2007.09.001} {\bibfield  {journal}
  {\bibinfo  {journal} {Nucl. Phys. A}\ }\textbf {\bibinfo {volume} {796}},\
  \bibinfo {pages} {41} (\bibinfo {year} {2007})},\ \Eprint
  {https://arxiv.org/abs/0708.0231} {arXiv:0708.0231 [hep-ph]} \BibitemShut
  {NoStop}%
\bibitem [{\citenamefont {Stasto}\ \emph {et~al.}(2012)\citenamefont {Stasto},
  \citenamefont {Xiao},\ and\ \citenamefont {Yuan}}]{Stasto:2011ru}%
  \BibitemOpen
  \bibfield  {author} {\bibinfo {author} {\bibfnamefont {A.}~\bibnamefont
  {Stasto}}, \bibinfo {author} {\bibfnamefont {B.-W.}\ \bibnamefont {Xiao}},\
  and\ \bibinfo {author} {\bibfnamefont {F.}~\bibnamefont {Yuan}},\ }\bibfield
  {title} {\bibinfo {title} {{Back-to-Back Correlations of Di-hadrons in dAu
  Collisions at RHIC}},\ }\href
  {https://doi.org/10.1016/j.physletb.2012.08.044} {\bibfield  {journal}
  {\bibinfo  {journal} {Phys. Lett. B}\ }\textbf {\bibinfo {volume} {716}},\
  \bibinfo {pages} {430} (\bibinfo {year} {2012})},\ \Eprint
  {https://arxiv.org/abs/1109.1817} {arXiv:1109.1817 [hep-ph]} \BibitemShut
  {NoStop}%
\bibitem [{\citenamefont {Kutak}\ and\ \citenamefont
  {Sapeta}(2012)}]{Kutak:2012rf}%
  \BibitemOpen
  \bibfield  {author} {\bibinfo {author} {\bibfnamefont {K.}~\bibnamefont
  {Kutak}}\ and\ \bibinfo {author} {\bibfnamefont {S.}~\bibnamefont {Sapeta}},\
  }\bibfield  {title} {\bibinfo {title} {{Gluon saturation in dijet production
  in p-Pb collisions at Large Hadron Collider}},\ }\href
  {https://doi.org/10.1103/PhysRevD.86.094043} {\bibfield  {journal} {\bibinfo
  {journal} {Phys. Rev. D}\ }\textbf {\bibinfo {volume} {86}},\ \bibinfo
  {pages} {094043} (\bibinfo {year} {2012})},\ \Eprint
  {https://arxiv.org/abs/1205.5035} {arXiv:1205.5035 [hep-ph]} \BibitemShut
  {NoStop}%
\bibitem [{\citenamefont {Albacete}\ \emph {et~al.}(2019)\citenamefont
  {Albacete}, \citenamefont {Giacalone}, \citenamefont {Marquet},\ and\
  \citenamefont {Matas}}]{Albacete:2018ruq}%
  \BibitemOpen
  \bibfield  {author} {\bibinfo {author} {\bibfnamefont {J.~L.}\ \bibnamefont
  {Albacete}}, \bibinfo {author} {\bibfnamefont {G.}~\bibnamefont {Giacalone}},
  \bibinfo {author} {\bibfnamefont {C.}~\bibnamefont {Marquet}},\ and\ \bibinfo
  {author} {\bibfnamefont {M.}~\bibnamefont {Matas}},\ }\bibfield  {title}
  {\bibinfo {title} {{Forward dihadron back-to-back correlations in $pA$
  collisions}},\ }\href {https://doi.org/10.1103/PhysRevD.99.014002} {\bibfield
   {journal} {\bibinfo  {journal} {Phys. Rev. D}\ }\textbf {\bibinfo {volume}
  {99}},\ \bibinfo {pages} {014002} (\bibinfo {year} {2019})},\ \Eprint
  {https://arxiv.org/abs/1805.05711} {arXiv:1805.05711 [hep-ph]} \BibitemShut
  {NoStop}%
\bibitem [{\citenamefont {Fujii}\ \emph {et~al.}(2020)\citenamefont {Fujii},
  \citenamefont {Marquet},\ and\ \citenamefont {Watanabe}}]{Fujii:2020bkl}%
  \BibitemOpen
  \bibfield  {author} {\bibinfo {author} {\bibfnamefont {H.}~\bibnamefont
  {Fujii}}, \bibinfo {author} {\bibfnamefont {C.}~\bibnamefont {Marquet}},\
  and\ \bibinfo {author} {\bibfnamefont {K.}~\bibnamefont {Watanabe}},\
  }\bibfield  {title} {\bibinfo {title} {{Comparison of improved TMD and CGC
  frameworks in forward quark dijet production}},\ }\href
  {https://doi.org/10.1007/JHEP12(2020)181} {\bibfield  {journal} {\bibinfo
  {journal} {JHEP}\ }\textbf {\bibinfo {volume} {12}},\ \bibinfo {pages}
  {181}},\ \Eprint {https://arxiv.org/abs/2006.16279} {arXiv:2006.16279
  [hep-ph]} \BibitemShut {NoStop}%
\bibitem [{\citenamefont {Caucal}\ \emph {et~al.}(2024)\citenamefont {Caucal},
  \citenamefont {Salazar}, \citenamefont {Schenke}, \citenamefont {Stebel},\
  and\ \citenamefont {Venugopalan}}]{Caucal:2023fsf}%
  \BibitemOpen
  \bibfield  {author} {\bibinfo {author} {\bibfnamefont {P.}~\bibnamefont
  {Caucal}}, \bibinfo {author} {\bibfnamefont {F.}~\bibnamefont {Salazar}},
  \bibinfo {author} {\bibfnamefont {B.}~\bibnamefont {Schenke}}, \bibinfo
  {author} {\bibfnamefont {T.}~\bibnamefont {Stebel}},\ and\ \bibinfo {author}
  {\bibfnamefont {R.}~\bibnamefont {Venugopalan}},\ }\bibfield  {title}
  {\bibinfo {title} {{Back-to-Back Inclusive Dijets in Deep Inelastic
  Scattering at Small x: Complete NLO Results and Predictions}},\ }\href
  {https://doi.org/10.1103/PhysRevLett.132.081902} {\bibfield  {journal}
  {\bibinfo  {journal} {Phys. Rev. Lett.}\ }\textbf {\bibinfo {volume} {132}},\
  \bibinfo {pages} {081902} (\bibinfo {year} {2024})},\ \Eprint
  {https://arxiv.org/abs/2308.00022} {arXiv:2308.00022 [hep-ph]} \BibitemShut
  {NoStop}%
\bibitem [{\citenamefont {Accardi}\ \emph {et~al.}(2016)\citenamefont {Accardi}
  \emph {et~al.}}]{Accardi:2012qut}%
  \BibitemOpen
  \bibfield  {author} {\bibinfo {author} {\bibfnamefont {A.}~\bibnamefont
  {Accardi}} \emph {et~al.},\ }\bibfield  {title} {\bibinfo {title} {{Electron
  Ion Collider: The Next QCD Frontier}: {Understanding the glue that binds us
  all}},\ }\href {https://doi.org/10.1140/epja/i2016-16268-9} {\bibfield
  {journal} {\bibinfo  {journal} {Eur. Phys. J. A}\ }\textbf {\bibinfo {volume}
  {52}},\ \bibinfo {pages} {268} (\bibinfo {year} {2016})},\ \Eprint
  {https://arxiv.org/abs/1212.1701} {arXiv:1212.1701 [nucl-ex]} \BibitemShut
  {NoStop}%
\bibitem [{\citenamefont {Zheng}\ \emph {et~al.}(2014)\citenamefont {Zheng},
  \citenamefont {Aschenauer}, \citenamefont {Lee},\ and\ \citenamefont
  {Xiao}}]{Zheng:2014vka}%
  \BibitemOpen
  \bibfield  {author} {\bibinfo {author} {\bibfnamefont {L.}~\bibnamefont
  {Zheng}}, \bibinfo {author} {\bibfnamefont {E.~C.}\ \bibnamefont
  {Aschenauer}}, \bibinfo {author} {\bibfnamefont {J.~H.}\ \bibnamefont
  {Lee}},\ and\ \bibinfo {author} {\bibfnamefont {B.-W.}\ \bibnamefont
  {Xiao}},\ }\bibfield  {title} {\bibinfo {title} {{Probing Gluon Saturation
  through Dihadron Correlations at an Electron-Ion Collider}},\ }\href
  {https://doi.org/10.1103/PhysRevD.89.074037} {\bibfield  {journal} {\bibinfo
  {journal} {Phys. Rev. D}\ }\textbf {\bibinfo {volume} {89}},\ \bibinfo
  {pages} {074037} (\bibinfo {year} {2014})},\ \Eprint
  {https://arxiv.org/abs/1403.2413} {arXiv:1403.2413 [hep-ph]} \BibitemShut
  {NoStop}%
\bibitem [{\citenamefont {Adams}\ \emph {et~al.}(2006)\citenamefont {Adams}
  \emph {et~al.}}]{STAR:2006dgg}%
  \BibitemOpen
  \bibfield  {author} {\bibinfo {author} {\bibfnamefont {J.}~\bibnamefont
  {Adams}} \emph {et~al.} (\bibinfo {collaboration} {STAR}),\ }\bibfield
  {title} {\bibinfo {title} {{Forward neutral pion production in p+p and d+Au
  collisions at $\sqrt{s_{NN}} = 200$~GeV}},\ }\href
  {https://doi.org/10.1103/PhysRevLett.97.152302} {\bibfield  {journal}
  {\bibinfo  {journal} {Phys. Rev. Lett.}\ }\textbf {\bibinfo {volume} {97}},\
  \bibinfo {pages} {152302} (\bibinfo {year} {2006})},\ \Eprint
  {https://arxiv.org/abs/nucl-ex/0602011} {arXiv:nucl-ex/0602011} \BibitemShut
  {NoStop}%
\bibitem [{\citenamefont {Braidot}(2010)}]{Braidot:2010zh}%
  \BibitemOpen
  \bibfield  {author} {\bibinfo {author} {\bibfnamefont {E.}~\bibnamefont
  {Braidot}} (\bibinfo {collaboration} {STAR}),\ }\bibfield  {title} {\bibinfo
  {title} {{Suppression of Forward Pion Correlations in d+Au Interactions at
  STAR}},\ }in\ \href@noop {} {\emph {\bibinfo {booktitle} {{45th Rencontres de
  Moriond on QCD and High Energy Interactions}}}}\ (\bibinfo {year} {2010})\
  pp.\ \bibinfo {pages} {355--338},\ \Eprint {https://arxiv.org/abs/1005.2378}
  {arXiv:1005.2378 [hep-ph]} \BibitemShut {NoStop}%
\bibitem [{\citenamefont {Adare}\ \emph {et~al.}(2011)\citenamefont {Adare}
  \emph {et~al.}}]{PHENIX:2011puq}%
  \BibitemOpen
  \bibfield  {author} {\bibinfo {author} {\bibfnamefont {A.}~\bibnamefont
  {Adare}} \emph {et~al.} (\bibinfo {collaboration} {PHENIX}),\ }\bibfield
  {title} {\bibinfo {title} {{Suppression of back-to-back hadron pairs at
  forward rapidity in $d+$Au Collisions at $\sqrt{s_{NN}}=200$ GeV}},\ }\href
  {https://doi.org/10.1103/PhysRevLett.107.172301} {\bibfield  {journal}
  {\bibinfo  {journal} {Phys. Rev. Lett.}\ }\textbf {\bibinfo {volume} {107}},\
  \bibinfo {pages} {172301} (\bibinfo {year} {2011})},\ \Eprint
  {https://arxiv.org/abs/1105.5112} {arXiv:1105.5112 [nucl-ex]} \BibitemShut
  {NoStop}%
\bibitem [{\citenamefont {Albacete}\ and\ \citenamefont
  {Marquet}(2011)}]{Albacete:2010rh}%
  \BibitemOpen
  \bibfield  {author} {\bibinfo {author} {\bibfnamefont {J.~L.}\ \bibnamefont
  {Albacete}}\ and\ \bibinfo {author} {\bibfnamefont {C.}~\bibnamefont
  {Marquet}},\ }\bibfield  {title} {\bibinfo {title} {{Single and double
  inclusive particle production in d+Au collisions at RHIC, leading twist and
  beyond}},\ }\href {https://doi.org/10.1016/j.nuclphysa.2010.09.014}
  {\bibfield  {journal} {\bibinfo  {journal} {Nucl. Phys. A}\ }\textbf
  {\bibinfo {volume} {854}},\ \bibinfo {pages} {154} (\bibinfo {year}
  {2011})},\ \Eprint {https://arxiv.org/abs/1009.3215} {arXiv:1009.3215
  [hep-ph]} \BibitemShut {NoStop}%
\bibitem [{\citenamefont {Chiu}(2011)}]{Chiu:2011ya}%
  \BibitemOpen
  \bibfield  {author} {\bibinfo {author} {\bibfnamefont {M.}~\bibnamefont
  {Chiu}} (\bibinfo {collaboration} {PHENIX}),\ }\bibfield  {title} {\bibinfo
  {title} {{Cold nuclear matter physics at forward rapidities from d+Au
  collisions in PHENIX}},\ }\href
  {https://doi.org/10.1088/0954-3899/38/12/124121} {\bibfield  {journal}
  {\bibinfo  {journal} {J. Phys. G}\ }\textbf {\bibinfo {volume} {38}},\
  \bibinfo {pages} {124121} (\bibinfo {year} {2011})},\ \Eprint
  {https://arxiv.org/abs/1109.2133} {arXiv:1109.2133 [nucl-ex]} \BibitemShut
  {NoStop}%
\bibitem [{\citenamefont {Chu}(2024)}]{ChuTalk}%
  \BibitemOpen
  \bibfield  {author} {\bibinfo {author} {\bibfnamefont {X.}~\bibnamefont
  {Chu}},\ }\href@noop {} {\bibinfo {title} {{Ongoing Efforts to Study Gluon
  Saturation at RHIC and at the LHC}}},\ \bibinfo {howpublished}
  {\url{https://indico.bnl.gov/event/20727/contributions/93759/}} (\bibinfo
  {year} {2024})\BibitemShut {NoStop}%
\bibitem [{\citenamefont {Kang}\ \emph {et~al.}(2012)\citenamefont {Kang},
  \citenamefont {Vitev},\ and\ \citenamefont {Xing}}]{Kang:2011bp}%
  \BibitemOpen
  \bibfield  {author} {\bibinfo {author} {\bibfnamefont {Z.-B.}\ \bibnamefont
  {Kang}}, \bibinfo {author} {\bibfnamefont {I.}~\bibnamefont {Vitev}},\ and\
  \bibinfo {author} {\bibfnamefont {H.}~\bibnamefont {Xing}},\ }\bibfield
  {title} {\bibinfo {title} {{Dihadron momentum imbalance and correlations in
  d+Au collisions}},\ }\href {https://doi.org/10.1103/PhysRevD.85.054024}
  {\bibfield  {journal} {\bibinfo  {journal} {Phys. Rev. D}\ }\textbf {\bibinfo
  {volume} {85}},\ \bibinfo {pages} {054024} (\bibinfo {year} {2012})},\
  \Eprint {https://arxiv.org/abs/1112.6021} {arXiv:1112.6021 [hep-ph]}
  \BibitemShut {NoStop}%
\bibitem [{\citenamefont {Armesto}\ \emph {et~al.}(2016)\citenamefont
  {Armesto}, \citenamefont {Paukkunen}, \citenamefont {Pen\'\i{}n},
  \citenamefont {Salgado},\ and\ \citenamefont {Zurita}}]{Armesto:2015lrg}%
  \BibitemOpen
  \bibfield  {author} {\bibinfo {author} {\bibfnamefont {N.}~\bibnamefont
  {Armesto}}, \bibinfo {author} {\bibfnamefont {H.}~\bibnamefont {Paukkunen}},
  \bibinfo {author} {\bibfnamefont {J.~M.}\ \bibnamefont {Pen\'\i{}n}},
  \bibinfo {author} {\bibfnamefont {C.~A.}\ \bibnamefont {Salgado}},\ and\
  \bibinfo {author} {\bibfnamefont {P.}~\bibnamefont {Zurita}},\ }\bibfield
  {title} {\bibinfo {title} {{An analysis of the impact of LHC Run I
  proton\textendash{}lead data on nuclear parton densities}},\ }\href
  {https://doi.org/10.1140/epjc/s10052-016-4078-9} {\bibfield  {journal}
  {\bibinfo  {journal} {Eur. Phys. J. C}\ }\textbf {\bibinfo {volume} {76}},\
  \bibinfo {pages} {218} (\bibinfo {year} {2016})},\ \Eprint
  {https://arxiv.org/abs/1512.01528} {arXiv:1512.01528 [hep-ph]} \BibitemShut
  {NoStop}%
\bibitem [{\citenamefont {Eskola}\ \emph {et~al.}(2019)\citenamefont {Eskola},
  \citenamefont {Paakkinen},\ and\ \citenamefont {Paukkunen}}]{Eskola:2019dui}%
  \BibitemOpen
  \bibfield  {author} {\bibinfo {author} {\bibfnamefont {K.~J.}\ \bibnamefont
  {Eskola}}, \bibinfo {author} {\bibfnamefont {P.}~\bibnamefont {Paakkinen}},\
  and\ \bibinfo {author} {\bibfnamefont {H.}~\bibnamefont {Paukkunen}},\
  }\bibfield  {title} {\bibinfo {title} {{Non-quadratic improved Hessian PDF
  reweighting and application to CMS dijet measurements at 5.02 TeV}},\ }\href
  {https://doi.org/10.1140/epjc/s10052-019-6982-2} {\bibfield  {journal}
  {\bibinfo  {journal} {Eur. Phys. J. C}\ }\textbf {\bibinfo {volume} {79}},\
  \bibinfo {pages} {511} (\bibinfo {year} {2019})},\ \Eprint
  {https://arxiv.org/abs/1903.09832} {arXiv:1903.09832 [hep-ph]} \BibitemShut
  {NoStop}%
\bibitem [{\citenamefont {Abdul~Khalek}\ \emph {et~al.}(2020)\citenamefont
  {Abdul~Khalek}, \citenamefont {Ethier}, \citenamefont {Rojo},\ and\
  \citenamefont {van Weelden}}]{AbdulKhalek:2020yuc}%
  \BibitemOpen
  \bibfield  {author} {\bibinfo {author} {\bibfnamefont {R.}~\bibnamefont
  {Abdul~Khalek}}, \bibinfo {author} {\bibfnamefont {J.~J.}\ \bibnamefont
  {Ethier}}, \bibinfo {author} {\bibfnamefont {J.}~\bibnamefont {Rojo}},\ and\
  \bibinfo {author} {\bibfnamefont {G.}~\bibnamefont {van Weelden}},\
  }\bibfield  {title} {\bibinfo {title} {{nNNPDF2.0: quark flavor separation in
  nuclei from LHC data}},\ }\href {https://doi.org/10.1007/JHEP09(2020)183}
  {\bibfield  {journal} {\bibinfo  {journal} {JHEP}\ }\textbf {\bibinfo
  {volume} {09}},\ \bibinfo {pages} {183}},\ \Eprint
  {https://arxiv.org/abs/2006.14629} {arXiv:2006.14629 [hep-ph]} \BibitemShut
  {NoStop}%
\bibitem [{\citenamefont {Kusina}\ \emph {et~al.}(2020)\citenamefont {Kusina}
  \emph {et~al.}}]{Kusina:2020lyz}%
  \BibitemOpen
  \bibfield  {author} {\bibinfo {author} {\bibfnamefont {A.}~\bibnamefont
  {Kusina}} \emph {et~al.},\ }\bibfield  {title} {\bibinfo {title} {{Impact of
  LHC vector boson production in heavy ion collisions on strange PDFs}},\
  }\href {https://doi.org/10.1140/epjc/s10052-020-08532-4} {\bibfield
  {journal} {\bibinfo  {journal} {Eur. Phys. J. C}\ }\textbf {\bibinfo {volume}
  {80}},\ \bibinfo {pages} {968} (\bibinfo {year} {2020})},\ \Eprint
  {https://arxiv.org/abs/2007.09100} {arXiv:2007.09100 [hep-ph]} \BibitemShut
  {NoStop}%
\bibitem [{\citenamefont {Eskola}\ \emph {et~al.}(2022)\citenamefont {Eskola},
  \citenamefont {Paakkinen}, \citenamefont {Paukkunen},\ and\ \citenamefont
  {Salgado}}]{Eskola:2021nhw}%
  \BibitemOpen
  \bibfield  {author} {\bibinfo {author} {\bibfnamefont {K.~J.}\ \bibnamefont
  {Eskola}}, \bibinfo {author} {\bibfnamefont {P.}~\bibnamefont {Paakkinen}},
  \bibinfo {author} {\bibfnamefont {H.}~\bibnamefont {Paukkunen}},\ and\
  \bibinfo {author} {\bibfnamefont {C.~A.}\ \bibnamefont {Salgado}},\
  }\bibfield  {title} {\bibinfo {title} {{EPPS21: a global QCD analysis of
  nuclear PDFs}},\ }\href {https://doi.org/10.1140/epjc/s10052-022-10359-0}
  {\bibfield  {journal} {\bibinfo  {journal} {Eur. Phys. J. C}\ }\textbf
  {\bibinfo {volume} {82}},\ \bibinfo {pages} {413} (\bibinfo {year} {2022})},\
  \Eprint {https://arxiv.org/abs/2112.12462} {arXiv:2112.12462 [hep-ph]}
  \BibitemShut {NoStop}%
\bibitem [{\citenamefont {Aaboud}\ \emph {et~al.}(2019)\citenamefont {Aaboud}
  \emph {et~al.}}]{ATLAS:2019jgo}%
  \BibitemOpen
  \bibfield  {author} {\bibinfo {author} {\bibfnamefont {M.}~\bibnamefont
  {Aaboud}} \emph {et~al.} (\bibinfo {collaboration} {ATLAS}),\ }\bibfield
  {title} {\bibinfo {title} {{Dijet azimuthal correlations and conditional
  yields in pp and p+Pb collisions at sNN=5.02TeV with the ATLAS detector}},\
  }\href {https://doi.org/10.1103/PhysRevC.100.034903} {\bibfield  {journal}
  {\bibinfo  {journal} {Phys. Rev. C}\ }\textbf {\bibinfo {volume} {100}},\
  \bibinfo {pages} {034903} (\bibinfo {year} {2019})},\ \Eprint
  {https://arxiv.org/abs/1901.10440} {arXiv:1901.10440 [nucl-ex]} \BibitemShut
  {NoStop}%
\bibitem [{\citenamefont {Abdallah}\ \emph {et~al.}(2022)\citenamefont
  {Abdallah} \emph {et~al.}}]{STAR:2021fgw}%
  \BibitemOpen
  \bibfield  {author} {\bibinfo {author} {\bibfnamefont {M.~S.}\ \bibnamefont
  {Abdallah}} \emph {et~al.} (\bibinfo {collaboration} {STAR}),\ }\bibfield
  {title} {\bibinfo {title} {{Evidence for Nonlinear Gluon Effects in QCD and
  Their Mass Number Dependence at STAR}},\ }\href
  {https://doi.org/10.1103/PhysRevLett.129.092501} {\bibfield  {journal}
  {\bibinfo  {journal} {Phys. Rev. Lett.}\ }\textbf {\bibinfo {volume} {129}},\
  \bibinfo {pages} {092501} (\bibinfo {year} {2022})},\ \Eprint
  {https://arxiv.org/abs/2111.10396} {arXiv:2111.10396 [nucl-ex]} \BibitemShut
  {NoStop}%
\bibitem [{\citenamefont {Strikman}\ and\ \citenamefont
  {Vogelsang}(2011)}]{Strikman:2010bg}%
  \BibitemOpen
  \bibfield  {author} {\bibinfo {author} {\bibfnamefont {M.}~\bibnamefont
  {Strikman}}\ and\ \bibinfo {author} {\bibfnamefont {W.}~\bibnamefont
  {Vogelsang}},\ }\bibfield  {title} {\bibinfo {title} {{Multiple parton
  interactions and forward double pion production in pp and dA scattering}},\
  }\href {https://doi.org/10.1103/PhysRevD.83.034029} {\bibfield  {journal}
  {\bibinfo  {journal} {Phys. Rev. D}\ }\textbf {\bibinfo {volume} {83}},\
  \bibinfo {pages} {034029} (\bibinfo {year} {2011})},\ \Eprint
  {https://arxiv.org/abs/1009.6123} {arXiv:1009.6123 [hep-ph]} \BibitemShut
  {NoStop}%
\bibitem [{\citenamefont {Adare}\ \emph {et~al.}(2014)\citenamefont {Adare}
  \emph {et~al.}}]{PHENIX:2013jxf}%
  \BibitemOpen
  \bibfield  {author} {\bibinfo {author} {\bibfnamefont {A.}~\bibnamefont
  {Adare}} \emph {et~al.} (\bibinfo {collaboration} {PHENIX}),\ }\bibfield
  {title} {\bibinfo {title} {{Centrality categorization for $R_{p(d)+A}$ in
  high-energy collisions}},\ }\href
  {https://doi.org/10.1103/PhysRevC.90.034902} {\bibfield  {journal} {\bibinfo
  {journal} {Phys. Rev. C}\ }\textbf {\bibinfo {volume} {90}},\ \bibinfo
  {pages} {034902} (\bibinfo {year} {2014})},\ \Eprint
  {https://arxiv.org/abs/1310.4793} {arXiv:1310.4793 [nucl-ex]} \BibitemShut
  {NoStop}%
\bibitem [{\citenamefont {Adam}\ \emph {et~al.}(2015)\citenamefont {Adam} \emph
  {et~al.}}]{ALICE:2014xsp}%
  \BibitemOpen
  \bibfield  {author} {\bibinfo {author} {\bibfnamefont {J.}~\bibnamefont
  {Adam}} \emph {et~al.} (\bibinfo {collaboration} {ALICE}),\ }\bibfield
  {title} {\bibinfo {title} {{Centrality dependence of particle production in
  p-Pb collisions at $\sqrt{s_{\rm NN} }$= 5.02 TeV}},\ }\href
  {https://doi.org/10.1103/PhysRevC.91.064905} {\bibfield  {journal} {\bibinfo
  {journal} {Phys. Rev. C}\ }\textbf {\bibinfo {volume} {91}},\ \bibinfo
  {pages} {064905} (\bibinfo {year} {2015})},\ \Eprint
  {https://arxiv.org/abs/1412.6828} {arXiv:1412.6828 [nucl-ex]} \BibitemShut
  {NoStop}%
\bibitem [{\citenamefont {Perepelitsa}(2024)}]{Perepelitsa:2024eik}%
  \BibitemOpen
  \bibfield  {author} {\bibinfo {author} {\bibfnamefont {D.~V.}\ \bibnamefont
  {Perepelitsa}},\ }\bibfield  {title} {\bibinfo {title} {{Contribution to
  differential $\pi^0$ and $\gamma_\mathrm{dir}$ modification in small systems
  from color fluctuation effects}},\ }\href
  {https://doi.org/10.1103/PhysRevC.110.L011901} {\bibfield  {journal}
  {\bibinfo  {journal} {Phys. Rev. C}\ }\textbf {\bibinfo {volume} {110}},\
  \bibinfo {pages} {L011901} (\bibinfo {year} {2024})},\ \Eprint
  {https://arxiv.org/abs/2404.17660} {arXiv:2404.17660 [nucl-th]} \BibitemShut
  {NoStop}%
\bibitem [{\citenamefont {van Hameren}\ \emph {et~al.}(2019)\citenamefont {van
  Hameren}, \citenamefont {Kotko}, \citenamefont {Kutak},\ and\ \citenamefont
  {Sapeta}}]{vanHameren:2019ysa}%
  \BibitemOpen
  \bibfield  {author} {\bibinfo {author} {\bibfnamefont {A.}~\bibnamefont {van
  Hameren}}, \bibinfo {author} {\bibfnamefont {P.}~\bibnamefont {Kotko}},
  \bibinfo {author} {\bibfnamefont {K.}~\bibnamefont {Kutak}},\ and\ \bibinfo
  {author} {\bibfnamefont {S.}~\bibnamefont {Sapeta}},\ }\bibfield  {title}
  {\bibinfo {title} {{Broadening and saturation effects in dijet azimuthal
  correlations in p-p and p-Pb collisions at $\mathbf{\sqrt{s}} = $ 5.02
  TeV}},\ }\href {https://doi.org/10.1016/j.physletb.2019.06.055} {\bibfield
  {journal} {\bibinfo  {journal} {Phys. Lett. B}\ }\textbf {\bibinfo {volume}
  {795}},\ \bibinfo {pages} {511} (\bibinfo {year} {2019})},\ \Eprint
  {https://arxiv.org/abs/1903.01361} {arXiv:1903.01361 [hep-ph]} \BibitemShut
  {NoStop}%
\bibitem [{\citenamefont {Bierlich}\ \emph {et~al.}(2022)\citenamefont
  {Bierlich} \emph {et~al.}}]{Bierlich:2022pfr}%
  \BibitemOpen
  \bibfield  {author} {\bibinfo {author} {\bibfnamefont {C.}~\bibnamefont
  {Bierlich}} \emph {et~al.},\ }\bibfield  {title} {\bibinfo {title} {{A
  comprehensive guide to the physics and usage of PYTHIA 8.3}},\ }\href
  {https://doi.org/10.21468/SciPostPhysCodeb.8} {\bibfield  {journal} {\bibinfo
   {journal} {SciPost Phys. Codeb.}\ }\textbf {\bibinfo {volume} {2022}},\
  \bibinfo {pages} {8} (\bibinfo {year} {2022})},\ \Eprint
  {https://arxiv.org/abs/2203.11601} {arXiv:2203.11601 [hep-ph]} \BibitemShut
  {NoStop}%
\bibitem [{\citenamefont {Duwent\"aster}\ \emph {et~al.}(2021)\citenamefont
  {Duwent\"aster}, \citenamefont {Husov\'a}, \citenamefont {Je\v{z}o},
  \citenamefont {Klasen}, \citenamefont {Kova\v{r}\'\i{}k}, \citenamefont
  {Kusina}, \citenamefont {Muzakka}, \citenamefont {Olness}, \citenamefont
  {Schienbein},\ and\ \citenamefont {Yu}}]{Duwentaster:2021ioo}%
  \BibitemOpen
  \bibfield  {author} {\bibinfo {author} {\bibfnamefont {P.}~\bibnamefont
  {Duwent\"aster}}, \bibinfo {author} {\bibfnamefont {L.~A.}\ \bibnamefont
  {Husov\'a}}, \bibinfo {author} {\bibfnamefont {T.}~\bibnamefont {Je\v{z}o}},
  \bibinfo {author} {\bibfnamefont {M.}~\bibnamefont {Klasen}}, \bibinfo
  {author} {\bibfnamefont {K.}~\bibnamefont {Kova\v{r}\'\i{}k}}, \bibinfo
  {author} {\bibfnamefont {A.}~\bibnamefont {Kusina}}, \bibinfo {author}
  {\bibfnamefont {K.~F.}\ \bibnamefont {Muzakka}}, \bibinfo {author}
  {\bibfnamefont {F.~I.}\ \bibnamefont {Olness}}, \bibinfo {author}
  {\bibfnamefont {I.}~\bibnamefont {Schienbein}},\ and\ \bibinfo {author}
  {\bibfnamefont {J.~Y.}\ \bibnamefont {Yu}},\ }\bibfield  {title} {\bibinfo
  {title} {{Impact of inclusive hadron production data on nuclear gluon
  PDFs}},\ }\href {https://doi.org/10.1103/PhysRevD.104.094005} {\bibfield
  {journal} {\bibinfo  {journal} {Phys. Rev. D}\ }\textbf {\bibinfo {volume}
  {104}},\ \bibinfo {pages} {094005} (\bibinfo {year} {2021})},\ \Eprint
  {https://arxiv.org/abs/2105.09873} {arXiv:2105.09873 [hep-ph]} \BibitemShut
  {NoStop}%
\bibitem [{\citenamefont {Abdul~Khalek}\ \emph {et~al.}(2022)\citenamefont
  {Abdul~Khalek}, \citenamefont {Gauld}, \citenamefont {Giani}, \citenamefont
  {Nocera}, \citenamefont {Rabemananjara},\ and\ \citenamefont
  {Rojo}}]{AbdulKhalek:2022fyi}%
  \BibitemOpen
  \bibfield  {author} {\bibinfo {author} {\bibfnamefont {R.}~\bibnamefont
  {Abdul~Khalek}}, \bibinfo {author} {\bibfnamefont {R.}~\bibnamefont {Gauld}},
  \bibinfo {author} {\bibfnamefont {T.}~\bibnamefont {Giani}}, \bibinfo
  {author} {\bibfnamefont {E.~R.}\ \bibnamefont {Nocera}}, \bibinfo {author}
  {\bibfnamefont {T.~R.}\ \bibnamefont {Rabemananjara}},\ and\ \bibinfo
  {author} {\bibfnamefont {J.}~\bibnamefont {Rojo}},\ }\bibfield  {title}
  {\bibinfo {title} {{nNNPDF3.0: evidence for a modified partonic structure in
  heavy nuclei}},\ }\href {https://doi.org/10.1140/epjc/s10052-022-10417-7}
  {\bibfield  {journal} {\bibinfo  {journal} {Eur. Phys. J. C}\ }\textbf
  {\bibinfo {volume} {82}},\ \bibinfo {pages} {507} (\bibinfo {year} {2022})},\
  \Eprint {https://arxiv.org/abs/2201.12363} {arXiv:2201.12363 [hep-ph]}
  \BibitemShut {NoStop}%
\bibitem [{\citenamefont {Buckley}\ \emph {et~al.}(2015)\citenamefont
  {Buckley}, \citenamefont {Ferrando}, \citenamefont {Lloyd}, \citenamefont
  {Nordstr\"om}, \citenamefont {Page}, \citenamefont {R\"ufenacht},
  \citenamefont {Sch\"onherr},\ and\ \citenamefont {Watt}}]{Buckley:2014ana}%
  \BibitemOpen
  \bibfield  {author} {\bibinfo {author} {\bibfnamefont {A.}~\bibnamefont
  {Buckley}}, \bibinfo {author} {\bibfnamefont {J.}~\bibnamefont {Ferrando}},
  \bibinfo {author} {\bibfnamefont {S.}~\bibnamefont {Lloyd}}, \bibinfo
  {author} {\bibfnamefont {K.}~\bibnamefont {Nordstr\"om}}, \bibinfo {author}
  {\bibfnamefont {B.}~\bibnamefont {Page}}, \bibinfo {author} {\bibfnamefont
  {M.}~\bibnamefont {R\"ufenacht}}, \bibinfo {author} {\bibfnamefont
  {M.}~\bibnamefont {Sch\"onherr}},\ and\ \bibinfo {author} {\bibfnamefont
  {G.}~\bibnamefont {Watt}},\ }\bibfield  {title} {\bibinfo {title} {{LHAPDF6:
  parton density access in the LHC precision era}},\ }\href
  {https://doi.org/10.1140/epjc/s10052-015-3318-8} {\bibfield  {journal}
  {\bibinfo  {journal} {Eur. Phys. J. C}\ }\textbf {\bibinfo {volume} {75}},\
  \bibinfo {pages} {132} (\bibinfo {year} {2015})},\ \Eprint
  {https://arxiv.org/abs/1412.7420} {arXiv:1412.7420 [hep-ph]} \BibitemShut
  {NoStop}%
\bibitem [{\citenamefont {Hou}\ \emph {et~al.}(2021)\citenamefont {Hou} \emph
  {et~al.}}]{Hou:2019efy}%
  \BibitemOpen
  \bibfield  {author} {\bibinfo {author} {\bibfnamefont {T.-J.}\ \bibnamefont
  {Hou}} \emph {et~al.},\ }\bibfield  {title} {\bibinfo {title} {{New CTEQ
  global analysis of quantum chromodynamics with high-precision data from the
  LHC}},\ }\href {https://doi.org/10.1103/PhysRevD.103.014013} {\bibfield
  {journal} {\bibinfo  {journal} {Phys. Rev. D}\ }\textbf {\bibinfo {volume}
  {103}},\ \bibinfo {pages} {014013} (\bibinfo {year} {2021})},\ \Eprint
  {https://arxiv.org/abs/1912.10053} {arXiv:1912.10053 [hep-ph]} \BibitemShut
  {NoStop}%
\bibitem [{Pri()}]{Private}%
  \BibitemOpen
  \href@noop {} {}\bibinfo {note} {Private communication, Xiaoxuan
  Chu}\BibitemShut {NoStop}%
\bibitem [{\citenamefont {Sjostrand}\ \emph {et~al.}(2006)\citenamefont
  {Sjostrand}, \citenamefont {Mrenna},\ and\ \citenamefont
  {Skands}}]{Sjostrand:2006za}%
  \BibitemOpen
  \bibfield  {author} {\bibinfo {author} {\bibfnamefont {T.}~\bibnamefont
  {Sjostrand}}, \bibinfo {author} {\bibfnamefont {S.}~\bibnamefont {Mrenna}},\
  and\ \bibinfo {author} {\bibfnamefont {P.~Z.}\ \bibnamefont {Skands}},\
  }\bibfield  {title} {\bibinfo {title} {{PYTHIA 6.4 Physics and Manual}},\
  }\href {https://doi.org/10.1088/1126-6708/2006/05/026} {\bibfield  {journal}
  {\bibinfo  {journal} {JHEP}\ }\textbf {\bibinfo {volume} {05}},\ \bibinfo
  {pages} {026}},\ \Eprint {https://arxiv.org/abs/hep-ph/0603175}
  {arXiv:hep-ph/0603175} \BibitemShut {NoStop}%
\bibitem [{\citenamefont {Skands}(2010)}]{Skands:2010ak}%
  \BibitemOpen
  \bibfield  {author} {\bibinfo {author} {\bibfnamefont {P.~Z.}\ \bibnamefont
  {Skands}},\ }\bibfield  {title} {\bibinfo {title} {{Tuning Monte Carlo
  Generators: The Perugia Tunes}},\ }\href
  {https://doi.org/10.1103/PhysRevD.82.074018} {\bibfield  {journal} {\bibinfo
  {journal} {Phys. Rev. D}\ }\textbf {\bibinfo {volume} {82}},\ \bibinfo
  {pages} {074018} (\bibinfo {year} {2010})},\ \Eprint
  {https://arxiv.org/abs/1005.3457} {arXiv:1005.3457 [hep-ph]} \BibitemShut
  {NoStop}%
\bibitem [{\citenamefont {Pumplin}\ \emph {et~al.}(2002)\citenamefont
  {Pumplin}, \citenamefont {Stump}, \citenamefont {Huston}, \citenamefont
  {Lai}, \citenamefont {Nadolsky},\ and\ \citenamefont
  {Tung}}]{Pumplin:2002vw}%
  \BibitemOpen
  \bibfield  {author} {\bibinfo {author} {\bibfnamefont {J.}~\bibnamefont
  {Pumplin}}, \bibinfo {author} {\bibfnamefont {D.~R.}\ \bibnamefont {Stump}},
  \bibinfo {author} {\bibfnamefont {J.}~\bibnamefont {Huston}}, \bibinfo
  {author} {\bibfnamefont {H.~L.}\ \bibnamefont {Lai}}, \bibinfo {author}
  {\bibfnamefont {P.~M.}\ \bibnamefont {Nadolsky}},\ and\ \bibinfo {author}
  {\bibfnamefont {W.~K.}\ \bibnamefont {Tung}},\ }\bibfield  {title} {\bibinfo
  {title} {{New generation of parton distributions with uncertainties from
  global QCD analysis}},\ }\href
  {https://doi.org/10.1088/1126-6708/2002/07/012} {\bibfield  {journal}
  {\bibinfo  {journal} {JHEP}\ }\textbf {\bibinfo {volume} {07}},\ \bibinfo
  {pages} {012}},\ \Eprint {https://arxiv.org/abs/hep-ph/0201195}
  {arXiv:hep-ph/0201195} \BibitemShut {NoStop}%
\bibitem [{\citenamefont {Boussarie}\ \emph {et~al.}(2017)\citenamefont
  {Boussarie}, \citenamefont {Grabovsky}, \citenamefont {Ivanov}, \citenamefont
  {Szymanowski},\ and\ \citenamefont {Wallon}}]{Boussarie:2016bkq}%
  \BibitemOpen
  \bibfield  {author} {\bibinfo {author} {\bibfnamefont {R.}~\bibnamefont
  {Boussarie}}, \bibinfo {author} {\bibfnamefont {A.~V.}\ \bibnamefont
  {Grabovsky}}, \bibinfo {author} {\bibfnamefont {D.~Y.}\ \bibnamefont
  {Ivanov}}, \bibinfo {author} {\bibfnamefont {L.}~\bibnamefont
  {Szymanowski}},\ and\ \bibinfo {author} {\bibfnamefont {S.}~\bibnamefont
  {Wallon}},\ }\bibfield  {title} {\bibinfo {title} {{Next-to-Leading Order
  Computation of Exclusive Diffractive Light Vector Meson Production in a
  Saturation Framework}},\ }\href
  {https://doi.org/10.1103/PhysRevLett.119.072002} {\bibfield  {journal}
  {\bibinfo  {journal} {Phys. Rev. Lett.}\ }\textbf {\bibinfo {volume} {119}},\
  \bibinfo {pages} {072002} (\bibinfo {year} {2017})},\ \Eprint
  {https://arxiv.org/abs/1612.08026} {arXiv:1612.08026 [hep-ph]} \BibitemShut
  {NoStop}%
\bibitem [{\citenamefont {Boussarie}\ \emph {et~al.}(2021)\citenamefont
  {Boussarie}, \citenamefont {M\"antysaari}, \citenamefont {Salazar},\ and\
  \citenamefont {Schenke}}]{Boussarie:2021ybe}%
  \BibitemOpen
  \bibfield  {author} {\bibinfo {author} {\bibfnamefont {R.}~\bibnamefont
  {Boussarie}}, \bibinfo {author} {\bibfnamefont {H.}~\bibnamefont
  {M\"antysaari}}, \bibinfo {author} {\bibfnamefont {F.}~\bibnamefont
  {Salazar}},\ and\ \bibinfo {author} {\bibfnamefont {B.}~\bibnamefont
  {Schenke}},\ }\bibfield  {title} {\bibinfo {title} {{The importance of
  kinematic twists and genuine saturation effects in dijet production at the
  Electron-Ion Collider}},\ }\href {https://doi.org/10.1007/JHEP09(2021)178}
  {\bibfield  {journal} {\bibinfo  {journal} {JHEP}\ }\textbf {\bibinfo
  {volume} {09}},\ \bibinfo {pages} {178}},\ \Eprint
  {https://arxiv.org/abs/2106.11301} {arXiv:2106.11301 [hep-ph]} \BibitemShut
  {NoStop}%
\bibitem [{\citenamefont {Armesto}\ \emph {et~al.}(2022)\citenamefont
  {Armesto}, \citenamefont {Lappi}, \citenamefont {M\"antysaari}, \citenamefont
  {Paukkunen},\ and\ \citenamefont {Tevio}}]{Armesto:2022mxy}%
  \BibitemOpen
  \bibfield  {author} {\bibinfo {author} {\bibfnamefont {N.}~\bibnamefont
  {Armesto}}, \bibinfo {author} {\bibfnamefont {T.}~\bibnamefont {Lappi}},
  \bibinfo {author} {\bibfnamefont {H.}~\bibnamefont {M\"antysaari}}, \bibinfo
  {author} {\bibfnamefont {H.}~\bibnamefont {Paukkunen}},\ and\ \bibinfo
  {author} {\bibfnamefont {M.}~\bibnamefont {Tevio}},\ }\bibfield  {title}
  {\bibinfo {title} {{Signatures of gluon saturation from structure-function
  measurements}},\ }\href {https://doi.org/10.1103/PhysRevD.105.114017}
  {\bibfield  {journal} {\bibinfo  {journal} {Phys. Rev. D}\ }\textbf {\bibinfo
  {volume} {105}},\ \bibinfo {pages} {114017} (\bibinfo {year} {2022})},\
  \Eprint {https://arxiv.org/abs/2203.05846} {arXiv:2203.05846 [hep-ph]}
  \BibitemShut {NoStop}%
\bibitem [{\citenamefont {Schenke}\ \emph {et~al.}(2024)\citenamefont
  {Schenke}, \citenamefont {M\"antysaari}, \citenamefont {Salazar},
  \citenamefont {Shen},\ and\ \citenamefont {Zhao}}]{Schenke:2024gnj}%
  \BibitemOpen
  \bibfield  {author} {\bibinfo {author} {\bibfnamefont {B.}~\bibnamefont
  {Schenke}}, \bibinfo {author} {\bibfnamefont {H.}~\bibnamefont
  {M\"antysaari}}, \bibinfo {author} {\bibfnamefont {F.}~\bibnamefont
  {Salazar}}, \bibinfo {author} {\bibfnamefont {C.}~\bibnamefont {Shen}},\ and\
  \bibinfo {author} {\bibfnamefont {W.}~\bibnamefont {Zhao}},\ }\bibfield
  {title} {\bibinfo {title} {{Vector meson production in ultraperipheral heavy
  ion collisions}},\ }in\ \href@noop {} {\emph {\bibinfo {booktitle} {{1st
  International Workshop on the physics of Ultra Peripheral Collisions}}}}\
  (\bibinfo {year} {2024})\ \Eprint {https://arxiv.org/abs/2404.10833}
  {arXiv:2404.10833 [nucl-th]} \BibitemShut {NoStop}%
\bibitem [{\citenamefont {Cheung}\ \emph {et~al.}(2024)\citenamefont {Cheung},
  \citenamefont {Kang}, \citenamefont {Salazar},\ and\ \citenamefont
  {Vogt}}]{Cheung:2024qvw}%
  \BibitemOpen
  \bibfield  {author} {\bibinfo {author} {\bibfnamefont {V.}~\bibnamefont
  {Cheung}}, \bibinfo {author} {\bibfnamefont {Z.-B.}\ \bibnamefont {Kang}},
  \bibinfo {author} {\bibfnamefont {F.}~\bibnamefont {Salazar}},\ and\ \bibinfo
  {author} {\bibfnamefont {R.}~\bibnamefont {Vogt}},\ }\bibfield  {title}
  {\bibinfo {title} {{Direct quarkonium production in DIS from a joint CGC and
  NRQCD framework}},\ }\href {https://doi.org/10.1103/PhysRevD.110.094039}
  {\bibfield  {journal} {\bibinfo  {journal} {Phys. Rev. D}\ }\textbf {\bibinfo
  {volume} {110}},\ \bibinfo {pages} {094039} (\bibinfo {year} {2024})},\
  \Eprint {https://arxiv.org/abs/2409.04080} {arXiv:2409.04080 [hep-ph]}
  \BibitemShut {NoStop}%
\bibitem [{\citenamefont {Beni\'c}\ \emph {et~al.}(2022)\citenamefont
  {Beni\'c}, \citenamefont {Garcia-Montero},\ and\ \citenamefont
  {Perkov}}]{Benic:2022ixp}%
  \BibitemOpen
  \bibfield  {author} {\bibinfo {author} {\bibfnamefont {S.}~\bibnamefont
  {Beni\'c}}, \bibinfo {author} {\bibfnamefont {O.}~\bibnamefont
  {Garcia-Montero}},\ and\ \bibinfo {author} {\bibfnamefont {A.}~\bibnamefont
  {Perkov}},\ }\bibfield  {title} {\bibinfo {title} {{Isolated photon-hadron
  production in high energy pp and pA collisions at RHIC and LHC}},\ }\href
  {https://doi.org/10.1103/PhysRevD.105.114052} {\bibfield  {journal} {\bibinfo
   {journal} {Phys. Rev. D}\ }\textbf {\bibinfo {volume} {105}},\ \bibinfo
  {pages} {114052} (\bibinfo {year} {2022})},\ \Eprint
  {https://arxiv.org/abs/2203.01685} {arXiv:2203.01685 [hep-ph]} \BibitemShut
  {NoStop}%
\end{thebibliography}%

\end{document}